\PassOptionsToPackage{dvipsnames,table}{xcolor}
\documentclass[acmsmall,screen,nonacm]{acmart}

\setcopyright{none}
\usepackage{tikz}
\usetikzlibrary{arrows.meta,positioning,shapes.geometric,fit,backgrounds,calc}
\usepackage[normalem]{ulem} %
\usepackage{listings}
\usepackage{tabularx}
\usepackage{makecell}
\usepackage{multirow}
\usepackage{arydshln} %
\usepackage{xspace}
\usepackage[capitalise]{cleveref}
\crefname{finding}{Finding}{Findings}
\Crefname{finding}{Finding}{Findings}

\lstdefinelanguage{Rust}{
  keywords={as,break,const,continue,crate,else,enum,extern,false,fn,for,if,impl,in,let,loop,match,mod,move,mut,pub,ref,return,self,Self,static,struct,super,trait,true,type,unsafe,use,where,while,dyn},
  sensitive=true,
  morecomment=[l]{//},
  morecomment=[s]{/*}{*/},
  morestring=[b]",
}
\lstdefinelanguage{diff}{
  morecomment=[f][\color{logerror}]{-},
  morecomment=[f][\color{darkgreen}]{+},
}
\lstdefinestyle{patch}{
  language=Rust,
  moredelim=**[l][\color{darkgreen}]{+},
  frame=tb,
  framerule=0.5pt,
  rulecolor=\color{gray!70},
  backgroundcolor=\color{gray!4},
  numbers=left,
  numberstyle=\tiny\color{gray},
  numbersep=6pt,
  xleftmargin=1.8em,
  framexleftmargin=1.8em,
  commentstyle=\itshape\color{gray},
}
\lstdefinestyle{prompt}{
  language={},
  basicstyle=\ttfamily\scriptsize,
  frame=single,
  framerule=0.4pt,
  rulecolor=\color{compilerblue!45},
  backgroundcolor=\color{compilerblue!4},
  framesep=5pt,
  xleftmargin=5pt,
  xrightmargin=5pt,
  breaklines=true,
  breakindent=0pt,
  moredelim=[il][\sffamily\itshape\color{compilerblue}]{\#\#\#\ },
  moredelim=[il][\sffamily\bfseries\color{compilerblue}]{\#\#\ },
  moredelim=[il][\sffamily\bfseries\small\color{compilerblue}]{\#\ },
  literate={[...]}{{\textcolor{gray}{[\ldots]}}}5,
  escapeinside={(*}{*)},
}
\newcommand{\pkey}[1]{\textbf{#1}}
\usepackage{pgfplots}
\pgfplotsset{compat=1.18}
\usepackage[most]{tcolorbox}

\newcommand{\lissue}[1]{\href{https://github.com/llvm/llvm-project/issues/#1}{L\##1}}
\newcommand{\rissue}[1]{\href{https://github.com/rust-lang/rust/issues/#1}{R\##1}}
\newcommand{\code}[1]{\texttt{#1}}

\newcommand{\rqone}{\textbf{RQ1}\xspace}
\newcommand{\rqtwo}{\textbf{RQ2}\xspace}
\newcommand{\rqthree}{\textbf{RQ3}\xspace}
\newcommand{\stepbadge}[1]{\smash{\tikz[baseline=(b.base)]\node[circle,draw=black,line width=0.3pt,fill=gray!25,inner sep=0.3pt,font=\sffamily\bfseries\fontsize{4}{4}\selectfont](b){#1};}}
\newcommand{\point}[1]{\smallskip\noindent\textbf{#1.}\xspace}
\newcolumntype{L}[1]{>{\raggedright\arraybackslash}p{#1}}
\newcolumntype{C}[1]{>{\centering\arraybackslash}p{#1}}
\newcolumntype{Y}{>{\raggedright\arraybackslash}X}
\newcolumntype{Z}{>{\centering\arraybackslash}X}

\definecolor{darkgreen}{RGB}{0,120,70}
\definecolor{compilerblue}{RGB}{20,103,139}
\definecolor{boxblue}{RGB}{205,216,229}
\definecolor{boxgreen}{RGB}{204,222,188}
\definecolor{boxpeach}{RGB}{239,220,210}
\definecolor{boxpink}{RGB}{235,200,233}
\definecolor{flowblue}{RGB}{222,235,245}
\definecolor{flowpink}{RGB}{246,211,241}
\definecolor{flowgray}{RGB}{210,210,210}
\definecolor{highlight}{RGB}{255,245,170}
\definecolor{loginfo}{RGB}{46,125,50}
\definecolor{logerror}{RGB}{198,40,40}

\newtcolorbox{findingbox}{colback=gray!6,colframe=gray!60,boxrule=0.4pt,arc=1pt,
  left=4pt,right=4pt,top=2pt,bottom=2pt,fontupper=\small}

\newtcolorbox{prbox}[1]{colback=white,colframe=gray!60,colbacktitle=gray!15,coltitle=black,boxrule=0.4pt,arc=1pt,
  left=4pt,right=4pt,top=2pt,bottom=2pt,fonttitle=\footnotesize\bfseries,fontupper=\footnotesize,parskip=3pt,title={#1}}
\newcommand{\lstnote}[1]{\quad{\color{compilerblue}\rmfamily\itshape\footnotesize #1}}
\newcounter{finding}
\newcommand{\finding}[2]{\refstepcounter{finding}\begin{findingbox}\textbf{Finding~\thefinding\ (#1).} #2\end{findingbox}}

\newcommand{\nRustcLabelled}{46}
\newcommand{\nRustcCandidates}{28}

\newcommand{\nLlvmCandidates}{124}
\newcommand{\nLlvmRust}{60}
\newcommand{\nLlvmFixes}{57}
\newcommand{\nDataset}{88}
\newcommand{\nProjects}{2,430}
\newcommand{\nSPProjects}{1,003}
\newcommand{\nJoltProjects}{90}

\newcommand{\nSPReproRustc}{15}
\newcommand{\nJoltReproRustc}{15}
\newcommand{\nSPRepro}{33}
\newcommand{\nJoltRepro}{31}
\newcommand{\nRQTwoProjects}{10}
\newcommand{\nRQThreeProjects}{20}

\newcommand{\nPatchLimit}{10}

\begin{document}

\title{Proving the Wrong Program: Miscompilations as a Threat to zkVM Guarantees}

\author{Hui Xuan Tan}
\affiliation{%
  \institution{Imperial College London}
  \country{United Kingdom}
}
\author{Stefanos Chaliasos}
\affiliation{%
  \institution{University of Athens}
  \country{Greece}
}
\affiliation{%
  \institution{ZKSecurity}
  \country{United States}
}
\author{Alastair F. Donaldson}
\affiliation{%
  \institution{Imperial College London}
  \country{United Kingdom}
}
\begin{abstract}
A zero-knowledge virtual machine (zkVM) lets an untrusted prover convince a verifier that a program, typically Rust compiled to RISC-V, ran correctly.
The proof covers the compiled binary, so a miscompilation can make a correct zkVM produce a valid proof of a computation that differs from the developer's source code.
We present the first empirical study of miscompilation as a threat to zkVM guarantees, on two production zkVMs, SP1 and Jolt.
We assemble \nDataset{} known rustc and LLVM miscompilations with Rust reproducers, using an LLM agent to recover Rust triggers for \nLlvmRust{} of \nLlvmCandidates{} LLVM bugs that were reported only as LLVM IR.
Many of the bugs carry over to zkVMs: 48 of the \nDataset{} make a guest commit a wrong output on SP1 and 46 on Jolt, including the same \nSPReproRustc{} rustc bugs on both.
Most bugs that do not reproduce never reach the faulty code on the zkVM's target, for example because they need an instruction-set extension.
Existing guest programs are rarely affected by accident.
Across 13 guests of \nRQTwoProjects{} popular SP1 projects, the buggy and the fixed compiler produce byte-identical binaries in 447 of the 552 guest-bug pairs that build, and the other 105 pairs have the same code or agree on every input we ran.
An attacker who contributes code to a guest, however, can trigger these bugs on purpose.
For about \$40 per project, an LLM agent found patches that look like ordinary optimisations, preserve the behaviour of the guest's source, and expose a known bug in 12 of 366 project-bug pairs, across 11 of \nRQThreeProjects{} projects.
In four cases the guest commits a wrong output, including an Ethereum client that rejects a valid block, and in the other eight it fails on a valid input.
Reviewing or formally verifying the zkVM or the guest's source cannot catch such a patch, because both are correct.
zkVM projects should therefore treat their compiler as part of their trusted base and track the known bugs of the toolchain they build with.
\end{abstract}



\maketitle

\section{Introduction}
\label{sec:intro}

Zero-knowledge virtual machines (zkVMs) let an untrusted prover produce a succinct proof that a program was executed correctly on specified inputs~\citep{zkvms-CHECKED, succinct00-CHECKED}.
Because checking a proof is typically much less expensive than re-executing the associated program~\citep{succinct-zkvm-CHECKED}, zkVMs now underpin application domains such as blockchains (for scalability~\cite{analysing-zk-rollups-CHECKED} and interoperability~\cite{zkBridge-CHECKED}), privacy-preserving machine learning and proofs of identity~\citep{lavin2026surveyapplicationszeroknowledgeproofs-CHECKED}.
In recent work on the quantum resources needed to break elliptic-curve cryptography, Google used SP1~\citep{SP1-CHECKED}, a popular zkVM, to back its claims without disclosing its improved quantum designs~\citep{GoogleEllipticCurves-CHECKED}.
Traditional approaches to zero-knowledge computation involve writing low-level arithmetic circuits that are error-prone~\cite{sok-snarks-CHECKED}.
The main benefit of zkVMs is that the program whose execution is proved---the \emph{guest}---can be written in an ordinary language such as Rust, and compiled with an existing compiler to the instruction set of the zkVM (usually RISC-V~\citep{argument24-CHECKED}).

The guarantees provided by zero-knowledge infrastructure rely on the infrastructure itself having been implemented correctly.
Infrastructure defects can have severe consequences.
For example, a September 2026 forged withdrawal led to the Payy protocol losing around \$1.9M~\citep{unchained-payy-report-CHECKED} despite them having written a correct zero-knowledge circuit: the problem turned out to be a bug in the underlying Noir/Barretenberg proof system~\citep{payy2026postmortem-CHECKED,smallgroup-bug-fix-CHECKED}.

In the context of zkVMs, a major trusted component is the compiler that compiles high-level source code to the zkVM's instruction set.
A zkVM proves that a given guest \emph{binary} was executed faithfully;
if the compiler \emph{miscompiles} the guest,
the prover will produce a valid proof of a computation that the developer did not actually write.
Despite substantial research on the security of zero-knowledge systems, examining hand-written circuits, proof systems and zkVM implementations~\citep{circom-vulns-CHECKED,sok-snarks-CHECKED,sok-understanding-zkvm-CHECKED},
and despite an Ethereum Foundation initiative towards formally verified zkVMs~\citep{verified-zkvms-CHECKED},
the security implications of miscompilation in zkVM guest compilers remain comparatively underexplored.
Miscompilation bugs are fairly common: between July 2024 and September 2026, the LLVM project developers fixed at least 124 miscompilations that could plausibly affect Rust code (\Cref{sec:bug-collection}), and the rustc developers have fixed dozens more miscompilations that affect rustc directly~\citep{rustc-compiler-bugs-CHECKED}.
Whether they matter in practice is less clear. \citet{Marcozzi-CHECKED} found that fuzzer-found and developer-reported Clang/LLVM bugs rarely changed the behaviour of Debian packages, suggesting that existing code seldom triggers known bugs.

A zkVM, however, operates in an adversarial setting, where the party that would benefit from a wrong result may also be able to influence the code that is compiled.
We consider an adversary who can contribute a patch to a guest program but does not control its build:
based on their knowledge of a latent compiler bug (either a publicly known bug that the maintainers' toolchain still contains, even if fixed upstream, or a bug that the adversary has discovered and kept secret)
they submit a benign-looking change that triggers a miscompilation in the maintainers' toolchain,
then supply inputs on which the compiled guest computes the wrong answer, and obtain a valid proof of it.

\textit{A motivating example.}
\Cref{lst:img-patch} shows a patch that an LLM agent wrote during our study (\Cref{sec:rq3-results}) for zk-img-editor, a small SP1 application that edits a signed photo and commits the result.
The patch adds a fast path that brightens or darkens the image in place, and leaves the remaining adjustments to the image library as before.
The pull-request message that the agent drafted presents the change as a way to avoid copying the image.
On every input, the patched source computes the same image as the original.
LLVM issue~\lissue{114181}\footnote{We write L\#$n$ for issue $n$ on the LLVM issue tracker and R\#$n$ for issue $n$ on the Rust issue tracker.} breaks this.
The guard in lines 5 and 6 should send an adjustment of $-128$ to the library.
LLVM's instruction combiner rewrites the negation in line~5 but keeps a \emph{no-signed-wrap} flag that no longer holds, so the buggy compiler evaluates the comparison in line~6 to false.
The fast path then runs with a magnitude of zero, and the guest commits the original pixels as the darkened image.
The LLVM~19.1.1 that ships with the official SP1~v4 toolchain has this bug.
As a result, SP1 would prove the wrong image, and a verifier would accept the proof.
From a code review point of view, a pull request contributing the patch of \Cref{lst:img-patch} \emph{does}
constitute a correct optimisation;
the reviewer or patch contributor could even use formal verification to prove correctness of their source-level change.
An attacker with advance knowledge of a miscompilation bug
(e.g.\ because they found it ahead of time using a compiler fuzzing tool and did not report it,
or because despite the bug being known and even fixed upstream, it still affects the toolchain associated with a zkVM)
could thus submit such a patch and keep a backdoor open for as long as the project builds with an affected toolchain.

\begin{figure}[t]
\begin{lstlisting}[style=patch,escapeinside={(*}{*)},caption={Patch that an LLM agent wrote for zk-img-editor, an SP1 application that checks a photo's camera signature, applies a list of edits and commits the edited image (\Cref{sec:rq3-results}). Added lines are green, and we omit the helper's call site. The patch keeps the behaviour of the source, but under LLVM~\lissue{114181} the comparison in line~6 is false for \code{delta} $=-128$, and the guest commits the image without darkening it. Below the patch is the pull-request message that the agent drafted for it.},label={lst:img-patch}]
+// Keep the common small adjustment in place; the library path handles values
+// outside this range and the two boundary cases.
+#[inline(never)] (*\lstnote{keep the helper a separate function}*)
+fn brighten_in_place_if_small_brightening(img: &mut RgbImage, delta: i8) -> bool {
+    let selected = if delta > 16 { -delta } else { delta }; (*\lstnote{\code{-delta} if \code{delta} $>16$, else \code{delta}}*)
+    if selected.wrapping_neg() == delta { (*\lstnote{true for \code{delta} $>16$, 0 and $-128$; the bug makes it false for $-128$}*)
+        return false; (*\lstnote{leave these values to the library}*)
+    }
+    let magnitude = if delta < 0 { (*\lstnote{how much to change each channel}*)
+        delta.checked_neg().unwrap_or_default() as u8 (*\lstnote{$|$\code{delta}$|$ when darkening, but 0 for $-128$}*)
+    } else {
+        delta as u8 (*\lstnote{\code{delta} when brightening}*)
+    };
+    for pixel in img.pixels_mut() { (*\lstnote{every pixel}*)
+        for channel in &mut pixel.0 { (*\lstnote{every RGB channel}*)
+            *channel = if delta < 0 {
+                channel.saturating_sub(magnitude) (*\lstnote{darken, stopping at 0}*)
+            } else {
+                channel.saturating_add(magnitude) (*\lstnote{brighten, stopping at 255}*)
+            };
+        }
+    }
+    true (*\lstnote{done here, skip the library}*)
+}
\end{lstlisting}

\begin{prbox}{Pull-request message drafted by the agent for \Cref{lst:img-patch} (abridged)}
\textbf{Apply common brightness adjustments in place.}
The transformation pipeline already owns a mutable RGB image. Today, even a small brightness adjustment goes through \code{imageops::brighten}, which produces a new image. This change handles the common bounded adjustments directly in the existing buffer, applying saturating arithmetic to each RGB channel. It is intended to avoid an intermediate image allocation and copy, particularly when several transformations run in sequence.

The fast path is deliberately limited to \code{-127..=-1} and \code{1..=16}. Zero, the \code{-128} signed boundary, larger positive adjustments, and values outside \code{i8} continue through the existing library implementation. [...] The input format, supported operations, and output format are unchanged; for fast-path values, saturating channel updates produce the same pixels as the existing implementation.

Correctness checks cover all 36,608 fast-path channel/value combinations, plus end-to-end comparisons against [...]
\end{prbox}
\end{figure}

In this paper we study the potential for miscompilation bugs to undermine the guarantees provided by zkVMs.
Our work is structured around the following research questions:

\begin{description}
\item[\rqone (Reproducibility)] Do known miscompilation bugs reproduce on zkVM targets, and if not, why not? (\Cref{sec:rq1-results})
\item[\rqtwo (Accidental impact)] Do existing zkVM guest programs trigger these bugs when built with a buggy compiler? (\Cref{sec:rq2-results})
\item[\rqthree (Adversarial impact)] Can a small patch that looks like an ordinary refactoring or optimisation make a guest program trigger a known bug? (\Cref{sec:rq3-results})\end{description}

To investigate these research questions, we first collected a pool of \nDataset{} miscompilation bugs affecting Rust programs,
drawing from the rustc and LLVM issue trackers.
In the case of LLVM miscompilations, which come with LLVM bytecode reproducers and cannot necessarily be reproduced when compiling from a higher-level language,
we used an LLM agent to identify Rust-level triggering programs for \nLlvmRust{} of the \nLlvmCandidates{} LLVM bugs.

For \rqone, we ran each reproducer on two zkVMs.
SP1~\citep{SP1-CHECKED} is the zkVM with the most open-source production projects at the time of writing (\Cref{sec:zkvm-selection}).
Jolt~\citep{jolt-paper-CHECKED} uses a different proof system and lets the developer choose the compiler for the guest.
For each bug, we built the reproducer for each zkVM with the buggy compiler and with the same compiler plus the upstream fix, and compared the two executions.
We found that many of the bugs carry over to zkVMs.
Of the \nDataset{} bugs, 48 make the guest commit a wrong output on SP1 and 46 on Jolt, and the zkVM would prove that output as readily as the correct one (\Cref{find:rq1}).
For most bugs that do not reproduce, the two compilers produce byte-identical guests, usually because the bug needs a feature that the zkVM's target lacks, such as an instruction-set extension (\Cref{find:rq1-why}).

For \rqtwo, we built 13 guest programs of \nRQTwoProjects{} open-source SP1 projects with the buggy and the fixed compiler of the bugs that reproduced on SP1 in \rqone.
Where the two binaries differed, we ran both builds on the same inputs from the project, such as test fixtures, and compared their outputs.
In 447 of the 552 pairs that built, the two compilers produce byte-identical guests, so no input can expose the bug.
In 9 more the guests differ only in layout, and the other 96 pairs agree on every input we ran (\Cref{find:rq2}).

For \rqthree, we took the attacker's view.
We gave an LLM agent \nRQThreeProjects{} SP1 projects, one at a time, and asked it to write patches that look like ordinary refactorings or optimisations, preserve the behaviour of the guest's source, and make the guest get affected by a known compiler bug.
For each patch, the agent built the guest with both compilers and searched for an input on which the two builds differ.
The agent succeeded for 12 of the 366 pairs of a project and a bug that it tried, in 11 of the \nRQThreeProjects{} projects, at a median cost of about \$40 per project.
In 4 of the 12, the guest commits a wrong output, among them an Ethereum client and the image editor of \Cref{lst:img-patch}.
In the other 8, the guest fails on a valid input, so the project can no longer prove that input (\Cref{find:rq3}).

In summary, our main contributions are:

\begin{itemize}

\item A thorough investigation into the potential for historic miscompilation bugs to affect the guarantees provided by zkVMs, via a study of \nDataset{} rustc and LLVM miscompilation bugs, showing that a large proportion of these bugs \emph{do} reproduce in a zkVM environment and thus have the \emph{potential} to affect the guarantees obtained for zkVM applications.

\item A study of the accidental impact of such bugs in practice, in which we build the 13 guests of \nRQTwoProjects{} popular SP1 projects with the buggy and the fixed compiler of each LLVM and rustc bug that reproduces on SP1. In 447 of the 552 pairs that built, the two compilers produce byte-identical guests, so no input could expose the bug, however much we tested. Only for the 96 pairs whose code differs does the result depend on our inputs, and their builds agree on every input we ran.

\item An investigation into whether an attacker can exploit miscompilation bugs on purpose, by contributing a patch that preserves the behaviour of a guest's source but makes the guest trigger a known bug. For about \$40 per project, an LLM agent found such patches for 12 of 366 project-bug pairs, in 11 of \nRQThreeProjects{} SP1 projects. In 4 the guest commits a wrong output.

\end{itemize}

\section{Background}
\label{sec:background}

\subsection{Zero-knowledge proofs and zkVMs}
\label{sec:zkvm-background}
\point{Zero-knowledge proofs} 
A zero-knowledge proof (ZKP) lets a \emph{prover} convince a \emph{verifier} that a statement is true without revealing anything beyond its truth~\citep{goldwasser1985knowledge-CHECKED}. 
For example, for a public function $F$ and public values $x$ and $y$, the prover shows that it knows an optionally private \emph{witness} $w$ such that $F(x;w) = y$. 
ZKPs are used both for privacy and for verifiable computation due to the succinctness of modern ZKPs~\cite{doyouneed-CHECKED}.
Two important properties of ZKP systems are \emph{soundness} and
\emph{completeness}. 
\emph{Soundness} means that a dishonest prover cannot convince the verifier of a false statement, except with negligible probability. 
\emph{Completeness} means that an honest prover can always convince the verifier of a true statement. 

\begin{figure}[t]
    \centering
    \begin{minipage}[c]{0.39\linewidth}
\begin{lstlisting}[language=Rust,frame=none,aboveskip=0pt,belowskip=0pt,basicstyle=\ttfamily\fontsize{5}{5.9}\selectfont,lineskip=-0.8pt]
// Guest: program/src/main.rs
#![no_main]
sp1_zkvm::entrypoint!(main);
fn main() {
  let x = sp1_zkvm::io::read::<i8>();
  sp1_zkvm::io::commit(&compute(x));
}
\end{lstlisting}
\vspace{1.5pt}\hrule\vspace{1.5pt}
\begin{lstlisting}[language={},frame=none,aboveskip=0pt,belowskip=0pt,basicstyle=\ttfamily\fontsize{5}{5.9}\selectfont,lineskip=-0.8pt]
(*@\stepbadge{1}@*) $ cd program && cargo prove build
\end{lstlisting}
\vspace{1.5pt}\hrule\vspace{1.5pt}
\begin{lstlisting}[language=Rust,frame=none,aboveskip=0pt,belowskip=0pt,basicstyle=\ttfamily\fontsize{5}{5.9}\selectfont,lineskip=-0.8pt]
// Host: script/src/main.rs
use sp1_sdk::{include_elf, ProverClient, SP1Stdin};
const ELF: &[u8] = include_elf!("program");
fn main() {
  let mut stdin = SP1Stdin::new();
  stdin.write(&-128i8);
  let client = ProverClient::from_env();
  client.execute(ELF, &stdin).run().unwrap();(*@\hfill\stepbadge{2}@*)
  let (pk, vk) = client.setup(ELF);
  let proof = client.prove(&pk, &stdin).run().unwrap();(*@\hfill\stepbadge{3}@*)
  client.verify(&proof, &vk).unwrap();(*@\hfill\stepbadge{4}@*)
}
\end{lstlisting}
    \end{minipage}\hfill
    \begin{minipage}[c]{0.59\linewidth}
    \includegraphics[width=\linewidth]{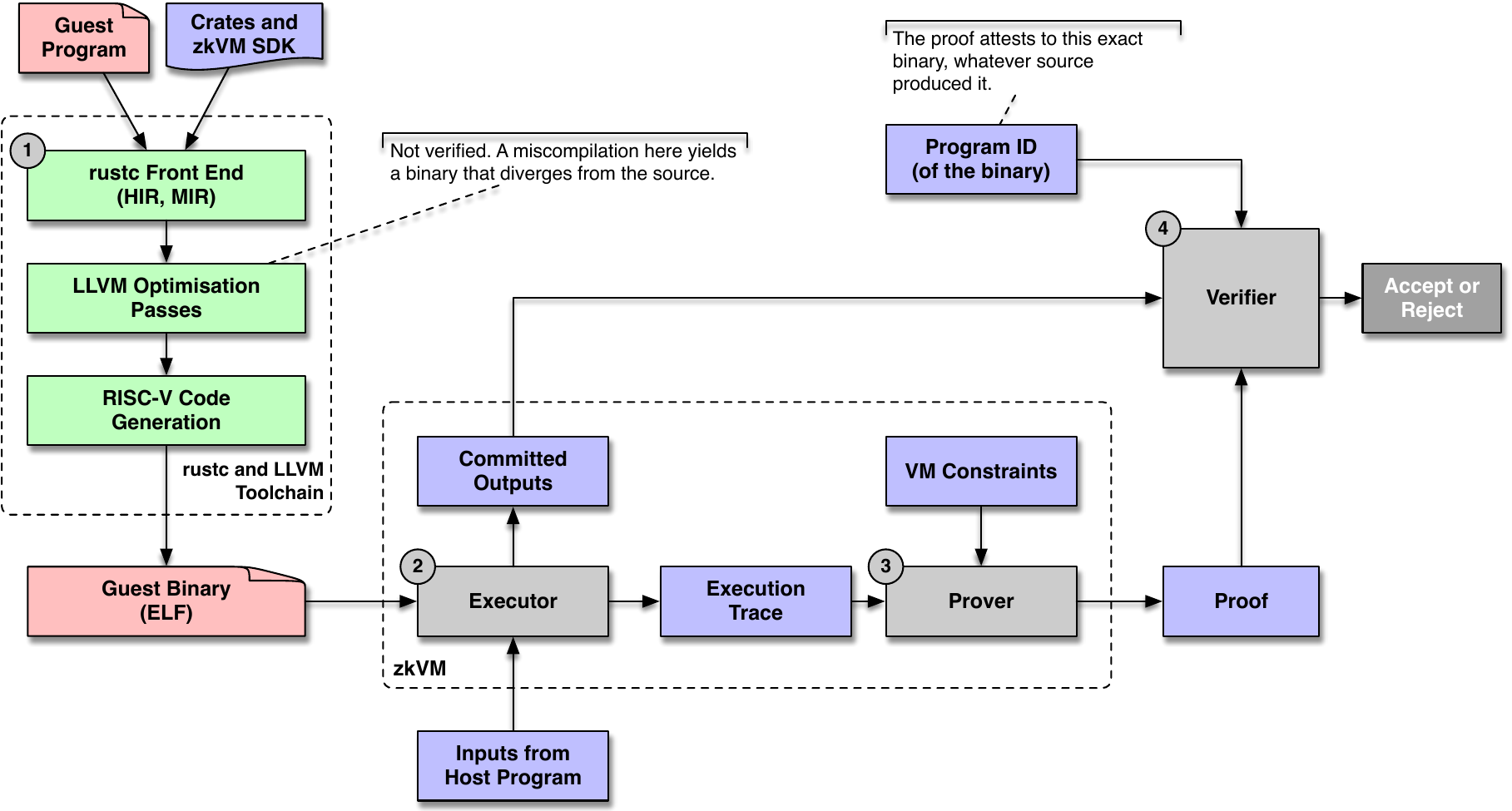}
    \end{minipage}
    \caption{Workflow of a zkVM application. Left: a minimal simplified SP1 application. Badges mark the code that performs each numbered step on the right. \code{vk}, the verifying key, plays the role of the program ID.}
    \label{fig:zkvm-workflow}
\end{figure}

\point{zkVMs} 
Traditionally, practitioners had to write low-level arithmetic circuits in DSLs like Circom, a task that requires expertise and is error-prone~\citep{sok-understanding-zkvm-CHECKED}.
A zkVM removes the need for users to write circuits.
Instead, its circuit encodes a processor for a fixed instruction set, most often a variant of RISC-V~\citep{sok-understanding-zkvm-CHECKED}, 
and so it can prove the execution of \emph{any} program expressed in that instruction set. 
\Cref{fig:zkvm-workflow} illustrates the workflow of a zkVM application. 
The developer writes the \emph{guest},
the program whose execution is proven, 
in an ordinary language, almost always Rust, 
and compiles it with an ordinary compiler to the zkVM's instruction set.
A \emph{host} program prepares the guest's inputs.
The zkVM's \emph{executor} runs the binary, and the guest \emph{commits} the values that it makes public, its outputs. 
The executor records a trace of every instruction and memory access, and the prover turns the trace into a proof. 
The verifier checks the proof against the public inputs, the committed outputs and an identifier of the binary, without running the program again. 
The proof therefore attests to the executor's run, and it is bound to the binary and not to the source.

\point{The zkVM as a compilation target} 
A zkVM target differs from an ordinary machine. 
It has no operating system, and the guest uses the zkVM's API/libraries for input,
output and \emph{precompiles}, 
which are dedicated circuits for costly operations such as hashing. 
The instruction set that a zkVM supports is typically small.
For example, the SP1 zkVM (discussed below)
does not support the floating-point or bit-manipulation extensions of RISC-V,
so a compiler bug that only affects code generation for such instructions cannot affect SP1.
When a check in the guest fails, 
the guest panics without committing anything, 
and when it performs an operation that the zkVM does not allow, 
the executor stops with a fault.
In both cases no proof of the expected output can be produced. 
Because the cost of a proof grows with the number of instructions that the guest executes,
real-world projects need to compile their guests with optimisations enabled for practical feasibility.
As a result, 
disabling optimisations is not a realistic defence against miscompilation bugs.

\point{SP1 and Jolt} 
SP1 is a popular RISC-V zkVM developed by Succinct~\citep{SP1-CHECKED}. 
Its guests are built with Succinct's fork of the Rust toolchain for a custom target. 
SP1~v4 ships LLVM~19.1.1 and targets \code{riscv32im},
and SP1~v6.8.1 targets \code{riscv64im}. 
A project picks its SP1 version, and with it the compiler, through its dependencies, but SP1 also allows a project to build the guest with another toolchain. 
Among current zkVMs, SP1 has the most open-source production projects (\Cref{sec:zkvm-selection}).

Jolt is a RISC-V zkVM developed by a16z~\citep{jolt-paper-CHECKED}. 
It proves the execution of instructions through lookup arguments and the sumcheck protocol~\citep{jolt-architecture}, 
targets \code{riscv64imac}, and allows the user to choose the guest's compiler.

\subsection{Compiling Rust with rustc and LLVM}
\label{sec:compiler-background}
The Rust compiler, rustc, compiles Rust code to machine code in two stages~\citep{rustc-docs}. 
The first is specific to Rust: rustc lowers the source through two intermediate representations of its own,
HIR and MIR,
checks types and borrows,
and optimises MIR.
The second is LLVM:
rustc translates MIR to LLVM IR, and then invokes LLVM which optimises the IR in a sequence of passes and generates code for the target architecture.
Passes that we refer to later in the paper include instruction combining (InstCombine), 
loop-invariant code motion (LICM), 
scalar evolution (SCEV) and interprocedural constant propagation (IPSCCP). 
Each rustc release bundles a specific LLVM version, and toolchains such as that of SP1 pin a rustc version and therefore also an LLVM version.

Broadly speaking, a compiler bug either crashes the compiler or makes it emit wrong code. 
The second kind, a \emph{miscompilation}, is the more dangerous, because the program compiles without complaint and misbehaves only at run time~\citep{Csmith}.
Our focus in this paper is on miscompilation bugs.
A Rust program can be miscompiled by a bug in rustc or in LLVM.

\subsection{Threat model}
\label{sec:threat-model}
We consider a correct zkVM (its proof system and verifier are both correct)
executing a zkVM application whose guest is built with a pinned toolchain that contains a miscompilation bug.
The bug may be publicly documented, possibly with a fix upstream that the pinned toolchain lacks, or it may be known only to an attacker.
A miscompilation can affect the guest in two ways.
(1) \emph{Accidentally}: existing guest code, including its dependencies, contains code that the buggy compiler miscompiles,
so the proven binary does not match the guest's source, without anyone acting maliciously.
\rqtwo measures how often this happens.
(2) \emph{Adversarially}: an attacker who knows of the bug contributes to the guest or to one of its dependencies. 
Following the pattern of recent supply-chain attacks~\citep{xz-backdoor}, 
the attacker can first build trust through useful contributions. 
It then proposes a change that looks like an ordinary refactoring or optimisation and preserves the behaviour of the source, 
so it passes review, 
but the pinned compiler miscompiles it. 
The attacker needs no access to the zkVM, 
the compiler or the prover, 
only knowledge of a bug in the project's toolchain. 
For a documented bug, this knowledge is public, i.e., the issue report typically includes a triggering test case, and a project that pins its toolchain stays exposed after the fix lands upstream. 
An undisclosed bug, which the attacker might find by compiler fuzzing~\citep{Csmith}, is more dangerous still, because there is neither a fix nor a report against which the project could check its toolchain. 
In both cases, the verifier accepts outputs that the source would never produce. 
\rqthree measures how easily such changes can be found.

Our study uses publicly documented bugs, because these are the only bugs we can observe and reproduce. 
Undisclosed bugs are part of the threat but beyond what we can measure. 
Our results for documented bugs therefore show what an undisclosed bug of the same kind could do.

\section{Methodology}
\label{sec:method}
\begin{figure}[t]
\centering
\includegraphics[width=0.85\linewidth]{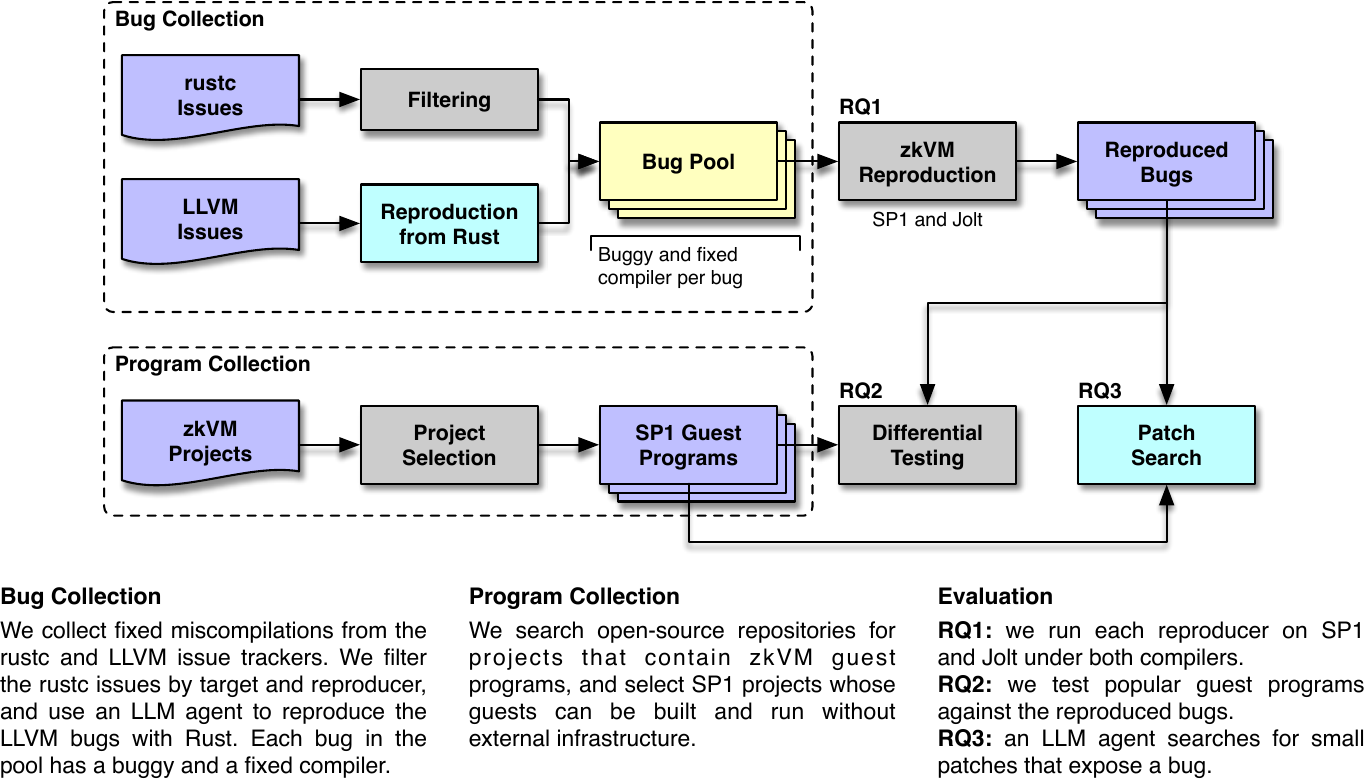}
\caption{Overview of the study. Bugs that reproduce on SP1 in \rqone are the input to \rqtwo and \rqthree, together with the selected SP1 guest programs. Steps in cyan are carried out by LLM agents and validated by humans.}
\label{fig:overview}
\end{figure}

First, we explain our process for obtaining a corpus of miscompilation bugs (\Cref{sec:bug-collection}).
Next, we explain how we selected the zkVMs to study and curated a dataset of real-world zkVM programs (\Cref{sec:zkvm-selection}).
Then, we describe the setup of experiments to answer each research question:
how we attempt to reproduce each bug on each zkVM target (\Cref{sec:rq1-method}),
how we test whether the bugs affect existing zkVM programs (\Cref{sec:rq2-method}),
and finally, how an attacker could craft a benign-looking patch that exposes a bug when the program is built with a buggy compiler (\Cref{sec:rq3-method}).
\Cref{fig:overview} gives an overview of our method.

\subsection{Bug collection}
\label{sec:bug-collection}
Our interest is in miscompilation bugs that could break the soundness of the proofs that a zkVM produces for a guest program.
zkVMs mainly support Rust, and most production projects write their guests in Rust.
We therefore require every bug in the study to meet three criteria.

\begin{description}
\item[C1 (Miscompilation).] The bug makes the compiler emit wrong code. Other kinds of compiler bug, such as crashes, are excluded.
\item[C2 (Rust reproducer).] The bug has a reproducer written in Rust, a known affected compiler version and an upstream fix.
\item[C3 (Supported version).] The affected version of rustc or LLVM is one that zkVMs support.
\end{description}

We started with miscompilations reported against versions of rustc that zkVMs support.
These bugs are specific to Rust, so they usually come with Rust reproducers.
However, rustc exhibits few historic bugs compared with older, more mature and complex compilers.
As discussed in \Cref{sec:background}, rustc relies on the LLVM compiler for most optimisations and for code generation, so that LLVM miscompilations have the potential to affect the compilation of Rust code.
We therefore also collected LLVM bugs, again from versions that zkVMs support.
Most LLVM bug reports come with a reproducer in LLVM-IR, whereas a zkVM takes a Rust program as input.
Writing a Rust program that triggers a given LLVM bug from an LLVM-IR reproducer is a time-consuming but largely mechanical task.
We did it by hand for three bugs, and then wrote our procedure down as instructions that an LLM agent follows within a fixed budget.
Below we describe the two sources, the agent's procedure and the resulting dataset.

\point{rustc bugs}
We searched the Rust compiler issue tracker for closed issues labelled \code{I-miscompile}, which returned \nRustcLabelled{} issues.
We removed 18 that affect only a non-RISC-V target, usually marked with an \code{O-} label such as \code{O-AVR}.\footnote{For example, \url{https://github.com/rust-lang/rust/issues/144076}}
We removed 8 more whose only reproducer is written in custom MIR (the intermediate representation used by Rustc).
Most of these were found by Rustlantis~\citep{Rustlantis-CHECKED}, which generates MIR directly to bypass the type and borrow checkers, so they have no Rust reproducer and fail \textit{C2}.
A thorough manual search added eight miscompilations that have a Rust reproducer but lack the label.
This gave \nRustcCandidates{} candidates, reported between July 2022 and February 2026. 
Eight of the \nRustcCandidates{} candidates are in fact LLVM bugs, whose faulty logic and fix lie in LLVM.
We keep them in this category because they were reported on the Rust issue tracker with a Rust reproducer, so unlike the LLVM bugs below they need no translation into Rust.

\point{LLVM bugs}
The LLVM issue tracker features far more miscompilations compared with that of rustc.
As the zkVMs we study compile guests with LLVM~19 or later (\Cref{sec:zkvm-selection}),
we compiled a list of \nLlvmCandidates{} LLVM miscompilations whose fixes landed between 24 July 2024 and 22 September 2026, to satisfy \textit{C3}.
Of these, 119 carry a miscompilation label, and 5 are unlabelled issues whose fixing commit describes a miscompilation.

\point{LLVM-to-Rust}
We ran Codex once per LLVM bug, with a wall-clock limit of four hours per run,
using model \code{gpt-5.6-sol} with high reasoning effort.
For each bug we prompted the agent to (a) build two compilers from the same rustc source, differing only in whether the LLVM compiler dependency has the fix applied,
and then (b) craft a Rust program whose output differs between the two.
Appendix~\ref{app:llvm-to-rust-prompt} gives the agent's instructions in condensed form.
An attempt ends in one of four outcomes (\Cref{tab:llvm-to-rust-outcomes}).
The agent may find a Rust program that diverges at run time between the two compilations (\emph{rust-runtime-confirmed}).
The other outcomes are:
the agent finds a Rust program that reaches the faulty IR transformation, but which does not trigger runtime divergence (\emph{rust-ir-confirmed});
the agent triggers the bug from Rust only with a modified pass pipeline or another diagnostic option (\emph{diagnostic-only});
the agent confirms the bug in LLVM but finds no Rust trigger (\emph{llvm-only-confirmed}).
We kept only the bugs with a runtime divergence through a Rust reproducer (\emph{rust-runtime-confirmed}), which left \nLlvmRust{} of the \nLlvmCandidates{} candidates.
A Rust program can trigger a different bug from the one under study, so we took three steps to tie each reproducer to its LLVM issue.
First, the agent's instructions ask for a program that triggers the given bug and no other, ideally with the same inputs and expected outputs as the LLVM reproducer.
Second, the two compilers differ only in the fix for that bug, and we checked that the compiler with the fix applied produces the correct result for every reproducer.
Third, we compared 5 reproducers by hand with their original LLVM issues, and each one matched.

\Cref{tab:llvm-to-rust-cost} summarises the cost of the \nLlvmCandidates{} candidate bugs.
The median run took 1.3 hours and used 28.1 million tokens.
Cached input accounts for 97.6\% of all tokens, because the agent resends its growing context at every step.
At the list API prices of \code{gpt-5.6-sol}, the median run costs \$15.21, and the \nLlvmCandidates{} runs cost \$2,013 in total.

\begin{table}[t]
\footnotesize
\begin{minipage}[t]{0.56\linewidth}
\setlength{\tabcolsep}{3.5pt}
\centering
\caption{Time, tokens and estimated cost per agent run of the LLVM-to-Rust step, over the \nLlvmCandidates{} completed runs. Cost uses the list API prices of \code{gpt-5.6-sol}: \$4.00, \$0.40 and \$20.00 per million uncached input, cached input and output tokens.}
\label{tab:llvm-to-rust-cost}
\begin{tabular}{@{}lrrrrr@{}}
\toprule
 & \textbf{Min} & \textbf{Median} & \textbf{Mean} & \textbf{Max} & \textbf{Total} \\
\midrule
Time (hours) & 0.13 & 1.30 & 1.41 & 3.71 & 175.4 \\
\midrule
Uncached input (millions) & 0.13 & 0.60 & 0.62 & 1.47 & 77.2 \\
Cached input (millions) & 3.53 & 27.47 & 29.42 & 62.77 & 3,648.1 \\
Output (millions) & 0.03 & 0.10 & 0.10 & 0.17 & 12.3 \\
All tokens (millions) & 3.70 & 28.06 & 30.14 & 64.28 & 3,737.5 \\
\midrule
Cost (USD) & 2.79 & 15.21 & 16.23 & 33.95 & 2,013.1 \\
\bottomrule
\end{tabular}
\end{minipage}\hfill
\begin{minipage}[t]{0.41\linewidth}
\centering
\caption{Outcome of the LLVM-to-Rust step for the \nLlvmCandidates{} candidates.}
\label{tab:llvm-to-rust-outcomes}
\begin{tabular}{@{}L{0.8\linewidth}r@{}}
\toprule
\textbf{Outcome} & \textbf{Bugs} \\
\midrule
Reproduced with a Rust program & \nLlvmRust{} \\
\addlinespace[3pt]
Faulty IR produced from a Rust program, without a runtime divergence & 19 \\
\addlinespace[3pt]
Reproduced with a Rust program, but only with custom flags or options & 13 \\
\addlinespace[3pt]
Reproduced only with the LLVM reproducer & 32 \\
\midrule
\textbf{Total} & \nLlvmCandidates{} \\
\bottomrule
\end{tabular}
\end{minipage}
\end{table}

\point{Dataset}
Our final dataset contains \nDataset{} bugs, each with a Rust reproducer: the \nRustcCandidates{} rustc candidates and the \nLlvmRust{} LLVM bugs reproduced from Rust.
These are the bugs that we try to reproduce on zkVMs in \rqone.
The \nLlvmRust{} LLVM bugs cover \nLlvmFixes{} distinct fixing commits, because three fixes each close two issues.
For these three pairs, the check against the fix cannot tell the two issues apart, because both issues share the same fixed compiler.
Here we rely on the agent matching the inputs and outputs of each issue's own LLVM reproducer.
The oldest bug in the dataset was reported in July 2022 and the newest in September 2026.

\point{Compiler pairs}
\label{sec:compiler-pairs}
Each bug in the dataset also comes with a buggy and a fixed compiler that differ only in the fix, so that we can attribute any difference in behaviour to the bug.
All three research questions use these pairs.
We build each pair from one rustc and LLVM revision.
When that revision still has the bug, it is the buggy compiler, and we apply the fix to get the fixed one.
When it already has the fix, it is the fixed compiler, and we undo the fix to get the buggy one.
The patch is the upstream fix where it applies cleanly, and otherwise a version adapted to the revision that changes as little as possible.
For the LLVM bugs, we reuse the pair that the agent built for LLVM-to-Rust.

\subsection{Selection of zkVMs and guest programs}
\label{sec:zkvm-selection}
In this work, we study two zkVMs, SP1 and Jolt, and a set of popular guest programs written for SP1.
To choose them, we first surveyed the open-source projects that contain zkVM guests.

\point{zkVMs}
We built a list of open-source projects with guest programs by searching GitHub for repositories that depend on a zkVM's guest crate or use its entry-point macro.
After removing forks and false positives, the list contains \nProjects{} projects.
We then classified the selected repos as production (projects either deployed on blockchains, or usable without modifications), work-in-progress (projects that are incomplete), benchmarks (repos containing benchmark payloads and comparisons between zkVMs), and tutorials (projects that are tutorials on how to use zkVMs).
Appendix~\ref{app:ecosystem} describes the procedure and gives the counts for each zkVM.

\Cref{fig:ecosystem} in Appendix~\ref{app:ecosystem} shows the seven zkVMs with the most projects.
RISC Zero has the most projects overall, 1,314 against \nSPProjects{} for SP1.
SP1 has the most projects classified as production, 149 against 108 for RISC Zero, and no other zkVM has more than 13.
We chose SP1 as our main zkVM because \rqtwo and \rqthree require guests that are deployed in practice, and SP1 has the most such guests.
Importantly, SP1 secures more than \$17B in the blockchain ecosystem versus \$400M from the second most deployed zkVM.\footnote{\url{https://l2beat.com/zk-catalog} as of October 2026}
A second reason is practical: our experiments require building every guest with compilers that we patch ourselves,
and SP1 allows a project to build its guests using a custom toolchain.
In contrast, RISC Zero manages its toolchain through its own installer, and we found no documented way to build a guest with a custom compiler.

We also wanted a second zkVM for \rqone, to assess the extent to which a bug's reproducibility is zkVM-dependent.
We picked Jolt because it has the third most projects (\nJoltProjects{}), it comes from a different team and uses a different proof system than SP1 (\Cref{sec:zkvm-background}), and (unlike RISC Zero) it lets the user choose which compiler to use for the guest.
Because Jolt is associated with only 4 production projects we only consider it for \rqone.
Both zkVMs compile guests with LLVM~19 or later: the SP1~v4 toolchain ships LLVM~19.1.1, and Jolt's dependencies require LLVM~19 or later.

\point{Guest programs}
\label{sec:programs}
The guest programs for \rqtwo and \rqthree come from the SP1 projects in our list.
We selected \nRQTwoProjects{} projects by applying two criteria:
(1) the guest must serve a real application, and we must be able to build and run it without external infrastructure;
(2) we must be able to provide inputs to the guest for testing.
The second criterion excludes,
for example,
Optimism's SP1 guests,\footnote{\url{https://github.com/ethereum-optimism/optimism/tree/develop/rust/kona/sp1}}
which need witnesses from a rollup node and from several layer-1 and layer-2 RPC endpoints, which public providers do not offer.
We ordered the projects meeting these criteria by popularity (GitHub stars) and selected:
\href{https://github.com/agglayer/agglayer}{agglayer},
\href{https://github.com/alpenlabs/alpen}{alpen},
\href{https://github.com/emberian/dregg}{dregg},
\href{https://github.com/lambdaclass/ethrex}{ethrex},
\href{https://github.com/grandinetech/grandine}{grandine},
\href{https://github.com/cosmos/ibc-contracts}{ibc-contracts},
\href{https://github.com/succinctlabs/rsp}{rsp},
\href{https://github.com/succinctlabs/sp1-helios}{sp1-helios},
\href{https://github.com/z0r0z/tacit}{tacit}, and
\href{https://github.com/tanujkhattar/zkp_ecc}{zkp\_ecc}.
Two of these (sp1-helios and ibc-contracts) contain multiple guest programs, and we keep all of them, leading to 13 guest programs in total.

\subsection{RQ1: Reproducing bugs on zkVM targets}
\label{sec:rq1-method}
\rqone asks whether the bugs in our dataset still show up when their reproducers run as zkVM guests.
A reproducer that diverges on a desktop target may behave differently here, because as discussed in \Cref{sec:zkvm-background} a zkVM compiles for a small RISC-V instruction set without an operating system.
We try each of the \nDataset{} bugs on both SP1 and Jolt.

\point{Compilers}
\label{sec:compilers}
We build each guest with the bug's compiler pair (\Cref{sec:compiler-pairs}).
The pair's rustc must be recent enough to build the zkVM's guest crates, which means Rust~1.86 or newer for SP1 and 1.88 or newer for Jolt.
When a rustc bug exists only in older releases, we re-introduce it into a newer rustc.
We exclude a bug when we cannot build such a pair that differs only in the fix.
For the LLVM bugs we use the same pair, flags, input and expected output on both zkVMs.

\point{Porting and execution}
We port each reproducer to a guest program that reads the reproducer's input and commits its output.
We build the guest with both compilers for the zkVM's target and run both builds in the zkVM's executor.
We do not generate any proofs as they are expensive to produce. 
Further, a proof attests to the execution that the executor performs, so a wrong output in the executor is the output that would be proven.
We use SP1 v6.8.1, which targets \code{riscv64im}, and Jolt at commit \code{bfcece09}, which targets \code{riscv64imac}.

\point{Outcomes}
A bug \emph{reproduces} on a zkVM when the buggy build commits a wrong output and the fixed build commits the expected one.
Any other outcome counts as not reproduced, and we record which of the following applies:
the two builds are byte-identical, so the bug does not change the guest;
the builds differ but commit the same output;
the buggy build traps or never terminates, so it has no output that could be proven;
the bug cannot be tested, because no clean compiler pair exists, the pair's rustc is too old to build the zkVM's guest crates, or the reproducer does not compile for the target.
For the bugs whose builds behave the same, we also give the likely reason, such as an instruction-set extension that the target lacks.

\subsection{RQ2: Testing existing guest programs}
\label{sec:rq2-method}
\rqtwo asks whether a bug that reproduces on SP1 changes the behaviour of real guest programs.
We follow the approach of \citet{Marcozzi-CHECKED}.
We build each program with the buggy and the fixed compiler, compare the two binaries, and run both where they differ.

\point{Pairs}
We use the \nRQTwoProjects{} SP1 projects of \Cref{sec:zkvm-selection}, which contain 13 guest programs.
We test every LLVM and rustc bug that reproduces on SP1 in \rqone{} (\nSPRepro{}) on every project.
This gives 624 (guest program, bug) pairs in total.

\point{Binary comparison}
For every pair we build the guest with the buggy and with the fixed compiler and compare the two binaries.
When they are byte-identical, the bug does not affect the guest and no input can make the two builds diverge.
We stop there for such a pair.
When the pair's compiler cannot build the guest, for example because the project's dependencies need a newer rustc, we record the pair as not buildable.

\point{Differential testing}
For the pairs whose binaries differ, we run both builds in SP1's executor on the same inputs, without proving, and compare what they commit.
zkVM guests rarely ship tests that run inside the zkVM, unlike the Debian packages of \citet{Marcozzi-CHECKED}.
We therefore take the inputs from the project where it provides them, such as test fixtures and official test vectors.
For projects where such inputs are unavailable we created sample inputs with the help of an LLM.
For every pair we record the build time of each compiler, whether the binaries are identical, the inputs, and the run time.
When two differing builds agree on every input under consideration, we do not know whether they are equivalent or our inputs missed the difference.
Still, it gives us an intuition if compiler bugs can easily affect zkVM guest programs.

\subsection{RQ3: Searching for patches that expose a bug}
\label{sec:rq3-method}
\rqthree asks how easily the attacker of \Cref{sec:threat-model} could succeed.
The attacker knows a bug in the project's compiler and wants to craft a patch to the guest that a reviewer would accept and that the buggy compiler miscompiles.
We give this task to an LLM agent.
Appendix~\ref{app:rq3-prompt} gives the agent's instructions in condensed form.

\point{Setup}
We run the agent on \nRQThreeProjects{} SP1 projects and give it access to the bugs from \rqone.
The projects are the \nRQTwoProjects{} of \rqtwo and 10 more from our list: automata-dcap-attestation, citrea, garbled-snark-verifier, hyli, op-succinct, prism, raiko, sp1-blobstream, sp1-storage-proof and zk-img-editor.
We used Codex with \code{gpt-6-sol} at medium effort, with a four-hour wall-clock timeout per project.

\point{Procedure}
The agent first builds the project's guests and runs them on a small set of inputs, to have a baseline.
For each bug it must then show that both compilers of the pair can build a real guest through the project's ordinary build.
Only then may it write patches.
It turns the bug's reproducer into a description of the operations, types, constants and control flow that the faulty transformation needs.
It searches the guest and the crates the guest uses for code that a small change would bring into that shape, and writes at most \nPatchLimit{} candidate patches per bug.
For each candidate it builds the guest with both compilers and checks, in the IR or the assembly, whether the buggy compiler applied the faulty transformation.
When it did, the agent works backwards from the affected code to the guest's input, to find an input that reaches it and makes the wrong value visible in the output.
It then runs both builds on that input, and on the project's existing tests if the first input shows nothing.
The agent only executes guests; it does not generate proofs.

\point{Patch rules}
A patch must be a plausible refactoring or optimisation of existing guest code that preserves semantics.
The agent may not paste the reproducer into unrelated code, add undefined behaviour, weaken an assertion, or change a test's expected value.
The two builds must use the same patched source, dependencies, build settings and input, and differ only in the compiler.

\point{Outcomes}
Each (project, bug) pair ends in one of four outcomes.
\emph{Success}: a valid patch for which the buggy and the fixed build produce different results on the same input, the fixed build agrees with the unpatched guest, and a replay from a clean state repeats the difference.
\emph{Success without test}: a valid patch for which the buggy compiler applied the faulty transformation, but no input showed a difference before the budget ended.
\emph{Partial}: a valid patch that built with both compilers but did not reach the faulty transformation.
\emph{Failure}: any other outcome, e.g.\ when no compiler pair could build the guest or where the agent found no plausible patch.

\subsection{Use of LLM agents}
\label{sec:llm-use}
Two of our experiments need more manual work than we could do by hand, so we gave them to LLM agents.
The first turns LLVM bugs into Rust reproducers, and the second searches for patches in \rqthree.
We wrote the instructions for both over a few trial runs, reading what the agent produced each time and changing the instructions where it went wrong (Appendix~\ref{app:llvm-to-rust-prompt} and Appendix~\ref{app:rq3-prompt}).
An author checked every agent result that the paper relies on.
The goal of this work is not to evaluate LLMs, and we make no claim about what they can or cannot do.
A different model, or different instructions, may do better or worse than ours did.
Outside these two experiments we used LLMs for small tasks only, such as preparing inputs for the guests of \rqtwo.

\point{Cost}
We ran both experiments under a ChatGPT Pro subscription, the 200 plan, which costs about 180 EUR per month.
We never reached the limits of the plan.
The costs we report are therefore what our tokens would cost at the list prices of the API, not what we paid (\Cref{tab:llvm-to-rust-cost} and \Cref{tab:rq3-cost}).

\point{LLVM-to-Rust}
Here the agent only gives us more bugs to test.
LLVM has a bug whether or not anyone finds a Rust program that triggers it, so the choice of model cannot change a bug.
It can change how many of the \nLlvmCandidates{} candidates end up in our dataset.
We keep a reproducer only if it diverges between the buggy and the fixed compiler when we replay it ourselves.

\point{Patch search}
In \rqthree, the agent searches for an adversarial patch that is semantically equivalent to the original source code but, with a given miscompilation, could lead to an exploit.
When our agent finds such a patch, the patch exists.
When it finds none, an attacker with more time and money may still be able to find one. 
Hence, we do not claim that our search is complete.

\subsection{Experimental setup}
\label{sec:setup}
We ran the LLVM-to-Rust step and the experiment of \rqthree on an Apple M1~Max machine with 10 cores and 64\,GB of RAM.
We validated the LLM-translated Rust reproducers for \rqone and ran the experiments of \rqtwo on a single AWS EC2 instance of type m7i.xlarge, which has 4 cores and 16GB of RAM and runs Ubuntu 24.04.

\section{Results}
\label{sec:results}
This section presents the main findings of our study. 
We first examine which bugs of our dataset reproduce on SP1 and Jolt, 
and why the others do not (RQ1, \Cref{sec:rq1-results}). 
Then, we test whether the bugs that reproduce change the behaviour of popular guest programs that are being used in production (RQ2, \Cref{sec:rq2-results}). 
Next, with the help of an LLM agent we examine whether we can find patches that expose a compiler bug in these programs that look like benign refactorings or optimisations.
Specifically, we discuss what the patches look like, 
how convincing pull request messages could be, 
and what the search costs are (RQ3, \Cref{sec:rq3-results}). 

\subsection{RQ1: Do known bugs reproduce on zkVM targets?}
\label{sec:rq1-results}

To learn whether a compiler bug can affect a zkVM, we first run its bug-triggering program through the zkVM's build pipeline.
One might expect this to always work, since the zkVM uses the same compiler, but often it does not.
As described in \Cref{sec:rq1-method}, we test the \nLlvmRust{} LLVM bugs that we reproduced from Rust and the \nRustcCandidates{} rustc bugs, each on SP1 and on Jolt.
We do not expect the results to differ much on other zkVMs with a RISC-V target, but a zkVM with another target, such as WebAssembly, could behave differently.
\Cref{tab:rq1} shows the outcomes. 

\textit{LLVM bugs.}
Of the \nLlvmRust{} LLVM bugs, \nSPRepro{} reproduce on SP1 (55.0\%) and \nJoltRepro{} on Jolt (51.7\%).
In each of them the buggy build commits a wrong output and the fixed build commits the expected one.
The two zkVMs agree almost everywhere.
Every bug that reproduces on Jolt also reproduces on SP1, and the two that reproduce only on SP1 could not be tested on Jolt, because their associated rustc version is too old to build Jolt's guest crates.

The bugs that reproduce come from many parts of LLVM.
Of the \nSPRepro{} that reproduce on SP1, 10 are in InstCombine, 5 in scalar evolution, 3 in ConstraintElimination, and 2 each in LICM and ValueTracking.
The other 11 are spread across 11 other components.
Most compiler pairs are also recent.
For 49 of the \nLlvmRust{} bugs the pair starts from one rustc revision that ships LLVM~23.1.1, and in 35 pairs this base already has the fix, so undoing it gives the buggy compiler.
These pairs show that, with a current compiler, Rust code can still reach the faulty logic on a zkVM target.

Four bugs change what the guest does without committing a wrong output.
In \lissue{142286} and \lissue{187472} the buggy build stops at a trap instruction that LLVM emitted for code it wrongly took to be unreachable.
In \lissue{116483} the buggy build never finishes.
SP1 ran it for an hour, and on Jolt the guest halted without an output.
In \lissue{124932} the buggy build runs out of memory on SP1.
We count these four as not reproduced, because no wrong output is committed.
Such a guest still fails on a valid input, but the failure cannot be used to prove a wrong result, so the impact is smaller.

\textit{rustc bugs.}
Of the \nRustcCandidates{} rustc bugs, the same \nSPReproRustc{} bugs reproduce on both SP1 and Jolt (53.6\%), where the buggy build commits a wrong output. 
These bugs come from various parts of the Rust compiler: three reside in rustc's MIR GVN pass, two come from rustc marking an argument \texttt{readonly} although the callee writes to it, and three are LLVM bugs reached through rustc (in Mem2Reg, in an implication between signed comparisons, and in machine copy propagation). The other seven are in seven different components.
Unlike the LLVM pairs, the rustc pairs do not share one base, because each bug lives in a different range of releases. SP1 needs Rust~1.86 or newer and Jolt Rust~1.88 or newer, so a bug that only older releases have was re-introduced into a newer compiler. For 6 of the 15 bugs on SP1 the buggy compiler is an official release or nightly. For the other nine we rebuild a newer rustc with the fix undone or by using the LLVM revision before the fix. Jolt needs one more such rebuild for \rissue{137867}, whose buggy release (Rust~1.86) is too old for Jolt's guest crates.

On SP1, five rustc bugs change what the guest does without committing a wrong output (three on Jolt). \rissue{136361} and \rissue{135997} are two reports of one GVN bug. There the buggy build casts a pointer to a type of a different size, and SP1 stops at the trap instruction emitted for it. In \rissue{152735} the guest calls through a vtable slot that rustc left empty. SP1 stops at a trap instruction, and Jolt at an unaligned store. In \rissue{142572} and \rissue{131834} the buggy build accesses an invalid address. SP1 rejects both runs, and on Jolt the first guest panics while the second accesses memory out of bounds. As with the LLVM bugs, we count these five as not reproduced, because no wrong output is committed.

\finding{RQ1}{\label{find:rq1}\emph{zkVMs can inherit the miscompilation bugs of the compilers they reuse.}
More than half of the LLVM and rustc bugs in our dataset reproduce on each zkVM.
Wherever a buggy compiler miscompiles a guest, the zkVM proves the wrong output as if it were correct.
This is worse than for an ordinary program, because the point of a zkVM is that others trust the output without running the program.
The bugs that reproduce are almost the same on SP1 and Jolt.
}

\begin{table}[t]
\centering
\footnotesize
\caption{\rqone: outcome of the \nDataset{} bugs of the dataset on SP1 and Jolt. A bug reproduces when the buggy build commits a wrong output and the fixed build the expected one.}
\label{tab:rq1}
\begin{tabular}{@{}lrrrr@{}}
\toprule
 & \multicolumn{2}{c}{\textbf{LLVM bugs}} & \multicolumn{2}{c}{\textbf{rustc bugs}} \\
\cmidrule(lr){2-3}\cmidrule(l){4-5}
\textbf{Outcome} & \textbf{SP1} & \textbf{Jolt} & \textbf{SP1} & \textbf{Jolt} \\
\midrule
\textbf{Reproduced} & \textbf{\nSPRepro{}} & \textbf{\nJoltRepro{}} & \textbf{\nSPReproRustc{}} & \textbf{\nJoltReproRustc{}} \\
\textbf{Not reproduced} & \textbf{21} & \textbf{20} & \textbf{7} & \textbf{5} \\
\quad The two builds are byte-identical & 15 & 15 & 0 & 0 \\
\quad The builds differ and commit the same output & 2 & 2 & 2 & 2 \\
\quad The buggy build traps or faults & 2 & 2 & 5 & 3 \\
\quad The buggy build does not terminate normally & 2 & 1 & 0 & 0 \\
\textbf{Not testable} & \textbf{6} & \textbf{9} & \textbf{6} & \textbf{8} \\
\quad No compiler pair that differs only in the fix & 3 & 3 & 0 & 0 \\
\quad rustc too old for the zkVM's guest crates & 2 & 5 & 5 & 7 \\
\quad Reproducer does not compile for the target or no MCVE & 1 & 1 & 1 & 1 \\
\bottomrule
\end{tabular}
\end{table}

\textit{Reasons for non-reproducibility.}
For 17 LLVM bugs the two builds behave the same on both zkVMs.
In 15 of these cases, the guests compiled with the buggy and fixed compilers are byte-identical, so the fix does not touch any code generated for the zkVM's target.
In 10 cases this is because the bug only triggers in the presence of features that the zkVM target lacks: 3 require the RISC-V Zbb or vector extensions; 3 require a specific instruction sequence only produced by x86 code generation; 2 require floating-point instructions, which the zkVM target only emulates in software; 1 requires tuning for a specific CPU; and 1 requires a 32-bit calling convention.
SP1~v4 targets \code{riscv32im} (\Cref{sec:zkvm-background}), so this last bug may still affect guests built with SP1~v4, which we did not test.
Of the other 5 cases, 2 never reach the faulty transformation, 2 sit behind a target check that RISC-V does not pass, and for the last the reason is undetermined. 

In the remaining 2 cases, \lissue{210137} and \lissue{216566}, the generated code does differ.
On a regular CPU the buggy program reads an invalid address and crashes.
In the zkVM that address is readable, so the read does not trap and the output matches the fixed build.

For 2 rustc bugs, \rissue{146163} and \rissue{112767}, the two builds differ but commit the same output on both zkVMs.
Unlike the byte-identical LLVM pairs, the difference in builds here is expected and bears no significance, because the two builds are compiled with different optimisation levels. 
\rissue{146163} is in LLVM's loop vectoriser, and the wrong code comes from a check that the vectoriser adds only when it groups memory accesses into vector operations.
Without vector instructions, which neither zkVM's target has, the loop is not vectorised and the check is never added.
\rissue{112767} is in instruction selection, which wrongly reuses the stack slot of a 128-bit argument passed in memory.
On x86-64 the reproducer's arguments fill all 6 argument registers, so the 128-bit value is passed on the stack.
The RISC-V calling convention has 8 argument registers, so the same value is passed in registers and there is no stack slot to reuse.

\finding{RQ1}{\label{find:rq1-why}
\emph{When a bug does not reproduce on a zkVM, it is usually because the zkVM's target never reaches the faulty code.}
In most such cases the buggy and the fixed compiler produce byte-identical guests, mostly because the bug requires something the zkVM's target lacks, such as an instruction-set extension, floating-point instructions, or code generation for another architecture like x86.
Other bugs cannot be tested at all, mostly because every compiler that has them is too old to build the zkVM's guests, or because their fix cannot be undone on a newer compiler.
So fewer compiler bugs affect zkVMs compared with ordinary targets, but---as per \Cref{find:rq1}---the bugs that do reproduce break the zkVM's guarantees.
}

\subsection{RQ2: Are existing guest programs affected?}
\label{sec:rq2-results}

\rqone shows that a miscompilation bug can affect the result of a zkVM when the zkVM is presented with a program crafted to trigger the bug.
The immediate next question is whether popular guest programs that are used in production are affected by such bugs when built with a buggy compiler. 
Here we ask whether the guests of \nRQTwoProjects{} popular zkVM projects contain code that a known bug could miscompile.
As described in \Cref{sec:rq2-method}, we took all \nSPRepro{} LLVM bugs and \nSPReproRustc{} rustc bugs that reproduce on SP1 for each of the \nRQTwoProjects{} projects, which gives 480 (project, bug) pairs.
Two projects have more than one guest, so the experiment covers 13 guests and 624 pairs of a guest and a bug.
\Cref{tab:rq2} presents the results of the experiment. 
Below we discuss each outcome and the main reason behind it.

\begin{table}[t]
\footnotesize
\begin{minipage}[t]{0.385\linewidth}
\centering
\caption{\rqtwo: comparison of the 624 pairs of a guest and a bug.}
\label{tab:rq2}
\begin{tabular}{@{}lrr@{}}
\toprule
\textbf{Result} & \textbf{Pairs} & \textbf{Share} \\
\midrule
Byte-identical builds & 447 & 71.6\% \\
Builds differ only in layout & 9 & 1.4\% \\
Builds differ & 96 & 15.4\% \\
Guest could not be built & 72 & 11.5\% \\
\midrule
\textbf{Total} & \textbf{624} & 100\% \\
\bottomrule
\end{tabular}
\end{minipage}\hfill
\begin{minipage}[t]{0.595\linewidth}
\centering
\caption{\rqtwo: the 96 pairs whose code differs, the inputs we ran on both builds, and the inputs on which the builds diverged (Div.).}
\label{tab:rq2-inputs}
\footnotesize
\setlength{\tabcolsep}{3pt}
\begin{tabular}{@{}lrrr@{\hspace{0.6em}};{2pt/2pt}>{\hspace{0.6em}}lrrr@{}}
\toprule
\textbf{Project} & \textbf{Pairs} & \textbf{Inputs} & \textbf{Div.} & \textbf{Project} & \textbf{Pairs} & \textbf{Inputs} & \textbf{Div.} \\
\midrule
ethrex   & 9 & 83 & 0 & sp1-helios    & 16 & 21   & 0 \\
dregg    & 10 & 68 & 0 & alpen         & 6 & 16   & 0 \\
grandine & 9 & 53 & 0 & agglayer      & 5 & 13   & 0 \\
tacit    & 6 & 23 & 0 & ibc-contracts & 17 & 1--7 & 0 \\
zkp\_ecc & 9 & 22 & 0 & rsp           & 9 & 4    & 0 \\
\bottomrule
\end{tabular}
\end{minipage}
\end{table}

\textit{Byte-identical builds.}
In 447 of the 552 pairs that built (81.0\%), the buggy and the fixed compiler produce the same guest, byte for byte. The fix changes nothing in these guests, so no input can make the two builds behave differently. A miscompilation needs code of a specific shape, such as a loop with a particular store, or a particular chain of comparisons. These guests contain no code of that shape, so the faulty transformation never fires and the fix has nothing to correct. The reproducers of \rqone are written to contain that structure, and real guests mostly do not. \citet{Marcozzi-CHECKED} observed a similar result for Clang and the Debian packages, where most fixes left most packages unchanged.

\textit{Guests that could not be built.}
For 72 (program, bug) pairs, the guest program could not be built: 12 guests with \lissue{124387}, 6 with \lissue{142518}, 4 each with \lissue{157238} and \lissue{178245}, 6 each with \rissue{111502}, \rissue{113407}, \rissue{127286}, \rissue{130853}, \rissue{131195}, \rissue{132353} and \rissue{137646}, and 4 with \rissue{135941}. All 12 bugs come with their own compiler pair on an older revision. For \lissue{124387} it is a rustc~1.86 nightly with LLVM~19.1.6, for \lissue{142518} a rustc~1.89 nightly with LLVM~20.1.5, for \lissue{157238} a rustc~1.91 nightly with LLVM~21.1.1, for \lissue{178245} a rustc~1.95 nightly with LLVM~23.0.0, for \rissue{135941} rustc~1.94.1 with LLVM~20.0.0, and for the other 7 aforementioned rustc bugs rustc~1.88 with LLVM~20.1.5. The dependencies of these guests need a newer rustc. This is the same limit that kept some bugs off Jolt in \rqone. A bug that lives only in an older compiler can only reach the projects that still build with that compiler.

\textit{Builds that differ.}
105 (program, bug) pairs produce different guests: 24 for LLVM bugs and 81 for rustc bugs (\Cref{tab:rq2-inputs}). For 9 of the rustc pairs, the two guests differ only in where code and data are placed, not in any instruction, leaving 96 pairs whose code actually differs.
The 72 rustc pairs come from 8 bugs. 
Five of them are in rustc's own optimisations on its mid-level intermediate representation (MIR), such as global value numbering and copy propagation, one is in compile-time evaluation, and two are in code generation.
The MIR passes run on every function of the guest and of the standard library, so a fix changes code often, whether or not the guest contains the code the bug needs.

The 24 LLVM pairs involve all 10 projects and 3 bugs: \lissue{60860}, \lissue{217052}, and \lissue{121890}.
\lissue{217052} changes all 13 guests, \lissue{60860} all but the two of sp1-helios and one of ibc-contracts, and \lissue{121890} only the guest of ethrex.
The compilers of all 3 pairs are  built from the same base, rustc~1.100.0-dev (commit \code{b37140ee}), which ships LLVM~23.1.1.
For \lissue{60860} and \lissue{217052}, the base is the buggy compiler and the fixed compiler is the base with the fix added.
\lissue{121890} is already fixed in the base, so the base is the fixed compiler, and the buggy compiler is the base with the fix undone.
The first two bugs are closely related.
Both are in LLVM's loop-invariant code motion (LICM) pass,
and both are fixed by the same commit.\footnote{LLVM commit \code{fbb86350}, ``Ignore AA tags on conditional stores in scalar promotion''.}
When a loop reads and writes the same memory location, LICM can keep the value in a register during the loop and store it back once after the loop.
That final store is executed even if the store inside the loop was conditional and never ran.
Before the fix, LICM used alias metadata on such a conditional store to decide that it did not touch the promoted location.
Alias metadata only holds for operations that run, so when the conditional store did not execute,
the promoted value could overwrite memory with a stale value.
The fix stops LICM from using the metadata in this case.
The third bug, \lissue{121890}, is in InstCombine and changes only the guest of ethrex.
When two addresses are computed from the same pointer and differ in a single index,
InstCombine can compare the two indices instead of the two addresses.
This is only correct if the address computation cannot wrap around,
which LLVM records as a flag on the computation.
Before the fix, InstCombine compared the indices even when the flag was missing,
so two indices that wrap around to the same address were reported as different.
The fix only folds the comparison when the flag is present.\footnote{LLVM commit \code{9725595f}, ``Check nowrap flags when folding comparison of GEPs with the same base pointer''.}

In line with \citet{Marcozzi-CHECKED}, the fact that a different binary is produced does not necessarily mean that a miscompilation has occurred.
For all three bugs, the fix turns off a transformation for code of a broad shape,
while only a small part of that code can be miscompiled.
The LICM fix makes LICM more conservative about every conditional store with alias metadata in a loop,
and Rust code often carries such metadata, to convey to LLVM that mutable references do not alias.
The fix therefore changes the code of any loop that has such a store,
whether or not the store can ever be skipped in the way the bug needs.
In the same way, the InstCombine fix changes every such comparison whose address computation lacks the flag,
although the result is only wrong when an index is large enough for the address to wrap around.
To learn whether the change matters, we ran the two builds of each pair in SP1's executor on the same inputs and compared their committed outputs.
For each guest we took the inputs from the project itself or from the data it is built to process:
its own tests and fixtures, the public Ethereum test vectors (ethrex and grandine), or recent Ethereum mainnet blocks (rsp) and light-chain data (sp1-helios) (\Cref{tab:rq2-inputs}).
The two builds committed the same output on every input, so we observed no difference in behaviour.
Together with the nature of the fix, this suggests that the bug was most likely never triggered. 

\finding{RQ2}{\label{find:rq2}\emph{The guests deployed today in popular projects are rarely affected by these known miscompilation bugs.} In 447 of the 552 pairs that built,
the buggy and the fixed compiler produce byte-identical guests, and in 9 more they differ only in layout, 
so the bug cannot affect them on any input.
The other 96 pairs come from two LLVM fixes and eight rustc bugs, and their two builds agree on every input we ran.}

\textit{Interpreting this result.}
For the 447 (program, bug) pairs that produced identical builds, we can confidently conclude that the bug does not affect the program, regardless of the input. The evidence is weaker for the 96 pairs whose code differs.
We ran them only on the projects' own inputs (1 to 83 per guest) and did not fuzz them, so we can only conclude that no divergence was observed on this set of inputs. 
That said, the fixes in these pairs disable a transformation for a broad class of code, so the differences most likely come from that wider effect rather than from miscompiled code.

There are two main limitations that apply here. First, we tested each project at one revision, but a guest can change with every commit. A later commit could introduce code with the shape that triggers the bug, although we expect this to be rare. Second, a project that builds with an old compiler is exposed to the bugs of that compiler, and we only tried the compilers of our pairs. 

\subsection{RQ3: Can a benign-looking contribution expose a known bug?}
\label{sec:rq3-results}

\rqone demonstrated that compiler miscompilation bugs can be reproduced in zkVMs,
while \rqtwo shows that popular guest programs rarely contain code that a known bug miscompiles.
\rqthree asks how hard it is to add such code on purpose.
Importantly, our insight is that a huge amount of resource and effort has been put into minimising soundness and completeness bugs in zkVMs through formal verification, 
but even a fully formally verified zkVM could be exploited via a compiler bug.
As discussed in \Cref{sec:intro}, it is common for attackers to invest significant funds and effort into performing an attack.
The threat model here is that an attacker might first spend time contributing valuable code to a zkVM guest project, becoming a trusted developer.
In the meantime, the attacker could scan the compiler used to compile guest programs for miscompilation bugs.
They would then be in a position to craft a contribution to the guest project---a refactoring or optimisation---that specifically introduces some code that triggers one or more miscompilations.

To assess the feasibility of this attack scenario, we give the task to an LLM agent (\Cref{sec:rq3-method}).
We expect a single autonomous agent to be orders of magnitude less effective than a real adversary with more resources.
The goal of this experiment is to produce patches that make a real guest trigger a known bug, and that also look like ordinary refactorings or optimisations that a maintainer would accept.
A patch that only needs to trigger the bug is easy to write, for example by pasting the bug's reproducer into the guest.
Such a patch would not survive review, so we count a patch only when it preserves the behaviour of the source and reads as a plausible change to the existing code.

The experiment was opportunistic.
We ran the agent on one project at a time, one after the other, and each run could read the reports of the earlier runs.
The agent chose which bugs to try on a project, wrote up to \nPatchLimit{} patches per bug and, for each patch, searched for an input on which the buggy and the fixed build behave differently (\Cref{sec:rq3-method}).
This gave the agent a lot of freedom in picking candidate bugs for each project and in how to exploit them.

Across the \nRQThreeProjects{} projects, the agent chose to try 366 project-bug pairs, skipping the bugs it judged irrelevant to a project.
\Cref{tab:rq3} summarises the campaign.
The agent found a \emph{successful} patch for 12 pairs (3.3\%), in 11 of the projects.
A success is a valid patch for which the buggy build produces a different result from the fixed build on the same input, and the fixed build agrees with the unpatched guest.
The successes come from only 5 bugs.
\lissue{219503} succeeded on seven projects,
\lissue{115149} on two,
and \lissue{114181},
\lissue{126012}
and \lissue{224832} on one each.
Because each run could read the reports of the earlier ones, later runs may have favoured bugs that had already succeeded, which would explain part of this concentration.
Another 17 pairs ended as \emph{partial}.
In these the agent wrote a valid patch that both compilers built, but the buggy compiler did not apply the faulty transformation to it.
The remaining 337 pairs failed, either because the bug's compiler pair could not build the guest or because the agent found no valid patch.
No pair ended as a \emph{success without test}.
Whenever a valid patch made the buggy compiler apply the faulty transformation, the agent also found an input on which the two builds differ.
An author reviewed every success.
The agent reported one more success, for ethrex with \lissue{126012}, which we rejected on review as a false positive (see below).

\begin{table}[t]
\footnotesize
\begin{minipage}[t]{0.52\linewidth}
\centering
\caption{\rqthree: statistics of the campaign, over 366 pairs of a project and a bug that the agent tried in \nRQThreeProjects{} projects.}
\label{tab:rq3}
\begin{tabular}{@{}lr@{}}
\toprule
\textbf{Statistic} & \textbf{Value} \\
\midrule
Pairs: success / success without test & 12 / 0 \\
Pairs: partial / failure & 17 / 337 \\
\midrule
Candidate patches written / valid & 149 / 138 \\
Patches reaching the faulty transformation & 21 \\
Compilations / guest executions & 1,157 / 1,050 \\
\midrule
Successful run, min / median / max (minutes) & 9 / 25 / 176 \\
\bottomrule
\end{tabular}
\end{minipage}\hfill
\begin{minipage}[t]{0.44\linewidth}
\centering
\caption{\rqthree: time per phase, from the activity records that the agent kept (23.8 hours in total).}
\label{tab:rq3-phases}
\begin{tabular}{@{}lrr@{}}
\toprule
\textbf{Phase} & \textbf{Hours} & \textbf{Share} \\
\midrule
Compiler pair & 11.4 & 48.0\% \\
Project setup and baseline & 5.0 & 21.2\% \\
Patch search & 2.7 & 11.3\% \\
Reporting & 1.8 & 7.7\% \\
Inputs, execution and replay & 1.6 & 6.5\% \\
Bug analysis & 1.3 & 5.3\% \\
\bottomrule
\end{tabular}
\end{minipage}
\end{table}

Most of the agent's time went into the compiler pair of each bug (\Cref{tab:rq3-phases}).
For each bug the agent provisioned the buggy and the fixed compiler, building them from one source revision with only the fix applied or reverted, and checked that both build the project's guest.
This took 48.0\% of the recorded time, setting up the project another 21.2\%, and the search for patches only 11.3\%.
For the projects of \rqtwo we could have shortened this step, since \rqtwo already shows which compiler pairs build which guests, but in this experiment we wanted to give the agent full freedom.

\begin{table}[t]
\centering
\footnotesize
\caption{\rqthree: the 12 successes, grouped by what the guest built with the buggy compiler does. \emph{Component} is the LLVM component that has the bug, and in every case the fixed build matches the unpatched guest. 
}
\label{tab:rq3-success}
\setlength{\tabcolsep}{5pt}
\begin{tabular}{@{}lllr@{}}
\toprule
\textbf{Project} & \textbf{Component} & \textbf{Code that the patch changes} & \textbf{Effect} \\
\midrule
\multicolumn{4}{@{}l}{\emph{The guest commits a wrong output (4)}} \\
\quad zk-img-editor & InstCombine & Brightness filter on RGB pixels & Wrong output \\
\quad zk-img-editor & InstCombine & Darkening filter, done in place & Wrong output \\
\quad ethrex & InstCombine & Common prefix of two trie paths & Wrong output \\
\quad sp1-blobstream & LoopLoadElim & Packing of the validator bitmap & Wrong output \\
\multicolumn{4}{@{}l}{\emph{The guest fails on a valid input (8)}} \\
\quad garbled-snark-verifier & SCEV & Check of an instance's wire slots & Panic \\
\quad alpen & IPSCCP & Merkle hashing of a checkpoint & Panic \\
\quad grandine & IPSCCP & Merkle tree of the state transition & Panic \\
\quad sp1-helios & IPSCCP & Hash of a beacon block header & Fault \\
\quad tacit & IPSCCP & Hash of a beacon block header & Fault \\
\quad ibc-contracts & IPSCCP & Merkle hash of a Tendermint header & Fault \\
\quad dregg & IPSCCP & BLAKE3 hashing of four chunks & Fault \\
\quad zkp\_ecc & IPSCCP & Parser of a qubit operand & Fault \\
\bottomrule
\end{tabular}
\end{table}

\textit{Overview of successful patches.}
\Cref{tab:rq3-success} lists the projects for which we managed to achieve a successful result, reporting the compiler component associated with the selected optimisation bug, a brief summary of the code change introduced by the patch, and the impact of the bug. The zk-img-editor project appears twice because we achieved two positive results for this project.

The four ``Wrong output'' results of \Cref{tab:rq3-success} are cases where the guest built with the buggy compiler finishes normally and commits a wrong output.
These are the most critical, because the zkVM would prove the wrong output.
The patches lead to: an Ethereum client rejecting a valid block (ethrex), a bridge committing a validator bitmap with 8 of its bytes set to zero (sp1-blobstream), and an image editor committing incorrect pixel values (zk-img-editor).
All four break soundness.
For the Ethereum client and the bridge, the results could be catastrophic.
The stakes of the image editor are low, since the bug only skips one darkening edit, but it shows what the proof gets wrong.
zk-img-editor proves the provenance of an image: its proof says that the committed image is a photo signed by a camera, with exactly the listed edits applied.
With the bug, the proof says that the photo was darkened when it was not, so a verifier accepts a false history for the image.

In the eight ``Panic'' and ``Fault'' cases, the buggy build panics or faults and commits nothing,
so there is no wrong output to prove.
An attacker can still use them, because they prevent a valid input from being proven.
This breaks completeness, and with it the liveness of the application: a rollup or a bridge that cannot prove a valid input stops making progress, and for a blockchain liveness is one of the main properties.
Seven of these cases come from one bug, \lissue{219503}, in which LLVM's interprocedural constant propagation drops the branch that ends a recursion.
The agent used the same idea on seven projects.
It specialised a recursive hash of a fixed number of fields or chunks,
and the buggy compiler removed the check that stops the recursion.

\textit{Reasons for failure to produce patches.}
There are two causes for the 337 failures.
In 117 cases the agent is not able to commence search because the compiler pair associated with the bug cannot build the project's guest.
A typical reason for this is that the bug requires an older compiler than the project's dependencies accept, the same limit we met in \rqone and \rqtwo.
On hyli no compiler pair built the guest at all, and on citrea, grandine and agglayer only 3, 6 and
7 did.
The compiler pair for bug \lissue{220904} could not build the guest associated with any project, and for bug \lissue{124387} only 6 projects could be built.
In the other 220 failure cases the compilers worked,
but the agent found no place in the guest where a plausible patch would produce the code needed to trigger the bug.
The 17 partial outcomes are valid patches that built with both compilers, but the buggy compiler did not apply the faulty transformation to them.
They come mostly from \lissue{126012} (5 pairs) and \lissue{219503} (4 pairs). 

\textit{Examples.}
Below we show two of the successful patches.
In the case of a successful patch, the agent also drafts the message for a pull request that would carry the patch,
and we quote the message associated with each example.
Recall that, as per the goal of the experiment, the agent was instructed to produce benign-looking patches together with convincing pull request messages.
If an attacker were following this approach in earnest,
they would likely go further and (a) provide some metrics for optimising patches to demonstrate their positive practical impact, and (b) work to hide the bug-triggering change inside a larger pull request to increase its chances of going unnoticed.

\emph{Example 1: An image filter (zk-img-editor, \lissue{114181}).} This is the patch of our motivating example in \Cref{sec:intro}.
The guest applies a list of transformations to an image and commits the result.
The patch in \Cref{lst:img-patch} adds a fast path that brightens or darkens the image in place for small adjustments and leaves the other values to the library.
For an adjustment of $-128$ the comparison is true and the request goes to the library, which darkens the image.
The buggy compiler evaluates the comparison to false.
The fast path then runs with a magnitude of zero and leaves the image as it was, and the guest commits pixels that were never darkened.

\emph{Example 2: An Ethereum client (ethrex, \lissue{115149}).}
In this project the guest executes a block against the state trie and commits whether the block is valid.
\Cref{lst:rq3-ethrex} changes the function that counts the common prefix of two trie paths.
It checks the first three bytes through a helper and falls back to the old comparison otherwise.
The helper walks a small array with a raw pointer until it reaches a zero byte.
The buggy compiler wrongly marks the pointer step in the loop as in bounds and so drops the store of the array's first byte.
The helper then reads a byte that was never written,
and the prefix comes out wrong.
With the fixed compiler,
the patched guest accepts a valid block from Ethereum's test suite,
as the unpatched guest does.
With the buggy compiler,
it commits that the block failed validation.
The other 42 bytes of its output are the same, and both executions exit normally.

\begin{figure}[t]
\begin{lstlisting}[style=patch,escapeinside={(*}{*)},caption={Patch for ethrex that exposes LLVM \lissue{115149}.},label={lst:rq3-ethrex}]
 fn count_common_prefix_scalar(a: &[u8], b: &[u8]) -> usize { (*\lstnote{length of the common prefix of two paths}*)
+    let n = a.len().min(b.len()); (*\lstnote{length of the shorter path}*)
+    // Most trie paths share a short initial run. Check three nibbles together before the comparison.
+    if n >= 3 && a[0] == b[0] { (*\lstnote{three nibbles, and the first matches}*)
+        let threshold = a[0].wrapping_add(1); (*\lstnote{one above the first nibble}*)
+        if threshold != 0 && threshold != u8::MAX { (*\lstnote{avoid overflow in the next lines}*)
+            let first_matches = a[1] == b[1]; (*\lstnote{do the second nibbles match?}*)
+            let first = threshold + u8::from(!first_matches); (*\lstnote{code: threshold, or one more on a mismatch}*)
+            let second = threshold + u8::from(a[2] != b[2]); (*\lstnote{same for the third nibbles}*)
+            let skip_first = if first_matches { 0 } else { -1 }; (*\lstnote{skip a code that matched}*)
+            if short_prefix_matches(first, second, threshold, skip_first) { (*\lstnote{second and third match?}*)
+                return 3 + a[3..n].iter().zip(b[3..n].iter()) (*\lstnote{count the rest}*)
+                    .take_while(|(x, y)| x == y).count();
+            }
+        }
+    }
     a.iter().zip(b.iter()).take_while(|(x, y)| x == y).count() (*\lstnote{old comparison, the fallback}*)
 }
+// A zero terminator lets the short common-prefix test stop without a second bounds branch. 
+#[inline(never)] (*\lstnote{keep the helper a separate function}*)
+fn short_prefix_matches(first: u8, second: u8, threshold: u8, skip_first: i8) -> bool {
+    if !(-1..=0).contains(&skip_first) || first == 0 || second == 0 { (*\lstnote{reject bad arguments}*)
+        return false;
+    }
+    let remaining = [first, second, 0]; (*\lstnote{codes, ended by a zero; the bug drops the store of \code{first}}*)
+    let start = (i32::from(skip_first) + 1) as isize; (*\lstnote{1 if the second nibbles matched, else 0}*)
+    let mut cursor = unsafe { remaining.as_ptr().offset(start) }; (*\lstnote{first code to read}*)
+    loop {
+        let code = unsafe { *cursor }; (*\lstnote{read a code}*)
+        if code == 0 { return true; } (*\lstnote{end reached, all matched}*)
+        if code > threshold { return false; } (*\lstnote{a mismatch}*)
+        cursor = unsafe { cursor.add(1) }; (*\lstnote{the bug marks this step as in bounds}*)
+    }
+}
\end{lstlisting}

\begin{prbox}{Pull-request message drafted by the agent for \Cref{lst:rq3-ethrex} (abridged)}
\textbf{Add a bounded three-nibble fast path to trie prefix counting.}
Specialize \code{count\_common\_prefix\_scalar} for trie paths that share at least three initial nibbles. The new path checks the first three bytes, then resumes the existing iterator comparison on the suffix when all three match. Short paths, first-byte mismatches, and threshold overflow cases continue through the original comparison.

The helper's three-byte array is fully initialized, and its signed start is checked before pointer use. Under correct compilation it returns true exactly when the two remaining prefix bytes match, so the returned prefix length is unchanged. No validation rule or guest output format changes. [...]
\end{prbox}
\end{figure}

\textit{Plausibility of the patches.}
The two patches have a few things in common. They are small: they add 33 and 44 lines, each in one function and one helper.
Each has a comment that gives a reason for performing this change that could convince a maintainer it is a useful change,
such as avoiding a copy or skipping a bounds check.
Each keeps the old code as the fallback for the inputs it does not handle,
which makes the change look like a careful optimisation.
The messages read like ordinary pull requests,
with a summary,
details of the cases that still take the old path,
and the tests that were run.
Those tests pass, because the source is correct and either the tests do not handle the cases where the output would be different,
or a non-buggy compiler is used.

A careful reviewer could still object,
especially if the change is proposed by a random developer,
or if it comes without benchmarks demonstrating optimisation gains.
With more instructions and a higher budget, an agent could likely make the patches even more realistic and accompany them with clear benchmarks showing performance gains.
The guard of \Cref{lst:img-patch} tests a range through a negation where two comparisons would do, and \Cref{lst:rq3-ethrex} needs \code{unsafe} for three bytes.
These are mainly objections of style.
They would not lead the reviewer to suspect that there may be a problem with the compiler, because nothing in the source is wrong.

In one instance, it was clear to us upon examining a success case that a reviewer would be likely to reject it.
For ethrex with \lissue{126012}, the agent rewrote the encoder of the guest's public output as a small state machine with a lagging index, and the buggy build panicked on a valid input.
After reviewing the patch, we observed that the state machine copies the structure of the bug's reproducer, and no reason is given for this change;
a direct pass over the fields would achieve equivalent results in a simpler way.
The patch also collects all fields in a vector and then copies them into a second one, which is not an improvement over the original code. 

\textit{Exploiting the guest.}
Exploiting a guest program after a malicious patch has been accepted may not be trivial,
and of course depends on the use of a compiler that is still prone to the bug in question.
However, we have noticed that zkVM projects and the zkVMs themselves update their compilation pipeline only infrequently.
Hence, an attacker could identify
a bug in a compiler version the zkVM has not yet adopted, get the patch merged, and exploit it once the toolchain is updated.

\textit{Cost of the search.}
The search we perform is cheap (\Cref{tab:rq3-cost}),
thus we expect a powerful attacker who invests significant resources would get substantially better results. 
Over the \nRQThreeProjects{} projects,
the median project used 157 million tokens and 2.5 hours of agent time,
which costs about \$40 at API prices. 
The most expensive project, grandine, cost \$76, and the \nRQThreeProjects{} projects together cost about \$803. Cached input makes up 98.2\% of all tokens. Successful runs took 9 minutes to 2.9 hours, with a median of 25 minutes.

\begin{table}[t]
\centering
\footnotesize
\caption{\rqthree: time, tokens and estimated cost per project, over the \nRQThreeProjects{} projects. Time is the time during which at least one agent of the project was active. Tokens are in millions. Cost uses the list API prices of \code{gpt-6-sol}: \$2.00, \$0.20 and \$10.00 per million uncached input, cached input and output tokens.}
\label{tab:rq3-cost}
\begin{tabular}{@{}lrrrrr@{}}
\toprule
 & \textbf{Min} & \textbf{Median} & \textbf{Mean} & \textbf{Max} & \textbf{Total} \\
\midrule
Time (hours) & 0.30 & 2.52 & 2.52 & 5.57 & 50.3 \\
\addlinespace
Uncached input tokens & 0.23 & 2.33 & 2.33 & 4.50 & 46.6 \\
Cached input tokens & 5.58 & 154.43 & 154.67 & 289.51 & 3,093.3 \\
Output tokens & 0.05 & 0.46 & 0.46 & 0.91 & 9.1 \\
All tokens & 5.86 & 157.45 & 157.45 & 294.93 & 3,149.0 \\
\addlinespace
\textbf{Cost (USD)} & \textbf{2.05} & \textbf{40.15} & \textbf{40.15} & \textbf{75.97} & \textbf{802.9} \\
\bottomrule
\end{tabular}
\end{table}

\finding{RQ3}{\label{find:rq3}\emph{A cheap agent can write plausible patches that expose known compiler bugs.} The agent found such a patch for 12 out of 366 (project, bug) pairs, in 11 of \nRQThreeProjects{} SP1 projects.
In 4 of the 12 cases this leads to the guest committing a wrong output,
including a case where an Ethereum client rejects a valid block.
In the other 8 the guest fails on a valid input.
The patches were generally well motivated, with convincing pull request messages.}

\section{Discussion}
\label{sec:discussion}

\subsection{Why existing guests rarely trigger a bug}
\label{sec:why-rare}
A miscompilation is triggered only by code that follows a specific pattern. 
The faulty optimisation must be reached, typically in a specific order, and its wrong result must then reach a value that the program commits.
The reproducers of \rqone are written to have that structure by definition (their raison d'\^etre is to expose their associated miscompilation bug),
but the guests of \rqtwo mostly do not, 
so in 94\% of the pairs that built, 
the fix has no effect on the binary that is produced. 
The zkVM's target further narrows the bugs that can affect a guest. 
Ten of the 15 LLVM bugs whose reproducers gave byte-identical builds in \rqone need a feature that the zkVM's target lacks, 
such as x86 code generation or the Zbb extension of RISC-V.

When a fix does change a guest, 
the change is usually broader than the bug. 
Both LLVM fixes behind the 24 differing LLVM pairs of \rqtwo
disable a transformation for a broad class of code patterns,
with only a subset of such patterns actually being miscompiled.
A binary difference thus provides weak evidence of a miscompilation.
\citet{Marcozzi-CHECKED} made the same observation for Clang, 
where many fixes changed the binaries of Debian packages without changing their behaviour.

\subsection{Implications}
\label{sec:implications}

\textit{Security of zkVMs and formal verification.}
A zkVM proof says only that a binary ran correctly on some inputs; anyone who relies on it is implicitly trusting that the binary conforms to its source.
The trusted base of a zkVM application therefore covers the proof system and its verifier, the zkVM's circuits, the guest's source and dependencies, and the toolchain that turns the source into the binary.
Most security work on zero-knowledge systems studies the first three~\citep{circom-vulns-CHECKED,sok-snarks-CHECKED,sok-understanding-zkvm-CHECKED,zkp-security-state-CHECKED}.
Our results show that the toolchain also deserves scrutiny: more than half of the LLVM bugs that we reproduced from Rust carry over to each zkVM, so a miscompilation can yield a valid proof of a result that the source program should not compute.
Much effort is being invested into verifying zkVMs, including an Ethereum Foundation initiative~\citep{verified-zkvms-CHECKED}, but a fully verified zkVM still proves exactly the binary it is given, miscompiled or not.
Formal verification of the guest's source does not help either,
because every malicious patch in \rqthree \emph{preserves} the behaviour of the source code.

The ideal solution would be a verified compiler such as CompCert~\citep{compcert-CHECKED}, but four things make this hard for zkVM guests.
First, no verified compiler covers Rust, which has no complete formal semantics yet.
Second, guests depend on many crates, including cryptographic libraries with \code{unsafe} code and the zkVM's own crates, all of which would have to go through the verified compiler.
Third, proof cost grows with the number of instructions executed, so projects depend on LLVM's optimisations, which is where most of our bugs sit (\Cref{sec:rq1-results}) and where verified compilers lag; a low optimisation level would make proofs infeasibly expensive.
Fourth, each zkVM has its own target and often its own toolchain fork, so each would need its own verified backend.
A lighter option is translation validation, which checks a single compilation rather than the whole compiler~\citep{translation-validation-CHECKED}; a zkVM guest is one fixed, rarely rebuilt binary, which makes it a good candidate.

\textit{zkVM developers.}
As discussed above, a zkVM's toolchain is part of the trusted base of the applications that are built on it. 
For example, the official SP1~v4 toolchain ships LLVM~19.1.1 without the fix for issue~\lissue{114181}, 
which miscompiles the patch in \Cref{lst:img-patch} on SP1's target.
To defend against miscompilation-based attacks,
zkVM teams can track the known miscompilations in the rustc and LLVM versions that they ship, 
backport the fixes that matter for their target, 
and publish which known bugs each toolchain version contains. 

\textit{zkVM guest program maintainers.}
A pull request that looks like an ordinary optimisation can introduce code designed to trigger a miscompilation (\rqthree).
Because the modified source is still correct,
this is likely to evade source code review.
As long as the project’s tests do not reach the affected input, or are currently run with a compiler that does not have the bug,
the malicious nature of the change will not be caught by testing either.
Two checks can help here.
The first is to build each change with multiple compilers:
the compiler associated with the pinned toolchain,
and one or more compilers that incorporate fixes for recent miscompilation bugs.
Most fixes leave a guest byte-identical (\rqtwo), 
so this will rarely raise an alarm, 
and when an alarm does occur it will point the reviewer to the source code whose associated binary representation has changed. 
The second is to look for unusual code patterns in a contribution,
but legitimate aggressive optimisations can feature unusual code patterns, making the actions of a careful attacker hard to detect.

\textit{The cost of an attack.}
\rqthree shows that an optimisation that preserves the behaviour of the source and triggers a known miscompilation is not hard to find.
The hard part for an attacker is to build enough reputation that an important project accepts its contributions.
The attackers described in \Cref{sec:threat-model} prepare a single attack for weeks or years, 
and such an attacker is likely to do better than our 12 successes in 366 pairs. 

\section{Threats to Validity}
\label{sec:threats}
We follow a standard methodology~\citep{feldt-validity-CHECKED} to identify validity threats, mitigating them where possible.

\point{Internal Validity}
A divergence that we attribute to a bug may have another cause.
The two compilers of a pair are built from the same rustc and LLVM revision and differ only in the fix; we exclude a bug when no such pair exists.
For 36 of the \nLlvmRust{} LLVM bugs the buggy compiler is a recent compiler with the fix undone (35 on the common LLVM~23.1.1 base of \Cref{sec:rq1-results} and \lissue{178245} on a rustc~1.95 nightly), sometimes with the patch adapted to the revision, so the pair may behave differently from the compiler that first had the bug.
We check each pair on the bug's reproducer, where the fixed compiler must produce the expected output, and in \rqthree the patched guest built with the fixed compiler must agree with the unpatched guest.
The agent that wrote the Rust reproducers was told to trigger the given LLVM bug and no other, keeping the inputs and expected outputs of the LLVM reproducer where possible; we checked five reproducers by hand against the original bug, and all five matched.
LLM agents can report outcomes they did not reach.
We keep an LLVM reproducer only when we can replay the claimed divergence ourselves, and an \rqthree success must repeat from a clean state; an author checked each success and rejected one.
The agent could see earlier runs, which may explain why 7 of its 12 successes use \lissue{219503}.
This affects how the successes spread over the bugs, not the effectiveness or credibility of any individual patch.
For the 96 \rqtwo pairs whose code differs, we ran between 1 and 83 inputs per guest without fuzzing, so we may have missed a divergence.
The 447 byte-identical pairs and the 9 that differ only in layout are unaffected, since no input can show a difference there.

\point{External Validity}
Our bugs may not represent all miscompilations that affect Rust.
The LLVM bugs come from fixes that landed between July 2024 and September 2026, and only the \nLlvmRust{} of \nLlvmCandidates{} candidates for which an agent found a Rust reproducer were tested; the rustc bugs leave out those with only a MIR reproducer.
For 49 LLVM bugs the pair is built on one recent rustc, so the results show that Rust code can trigger a bug on a zkVM target, not which bugs a zkVM release ships.
SP1 and Jolt are both RISC-V machines, and we run \rqtwo and \rqthree on SP1 only.
The two agree in \rqone, which suggests that the target matters more than the zkVM, but a zkVM with another instruction set may differ.
Our projects are popular open-source projects tested at one revision, excluding those that need infrastructure we could not provide, such as Optimism.
The \rqthree results hold for one model, one set of instructions and a few hours per project.
A success shows that a patch exists, but the success rate says little about how a more powerful attacker would perform.

\point{Construct Validity}
We did not generate proofs.
A zkVM proves the execution that its executor performs, so the output the executor commits is the output a proof would attest to; in early experiments we did produce proofs, and they followed the executor every time.
We dropped them when scaling up our study because they are so expensive to produce.
We checked that each successful patch preserves the behaviour of the source via tests and manual review, and judged its plausibility ourselves; that judgement rejected one patch, and a maintainer might reject others.
For ethical reasons we did not submit any patches, so \rqthree measures whether such a patch can be found, not how often one would be merged in practice.
Last, the costs we report are what our tokens would cost through the API, not what we paid (\Cref{sec:llm-use}).

\section{Related Work}
\label{sec:related}

\point{Compiler bugs and their impact}
There is a long line of work on finding compiler bugs.
Csmith~\citep{Csmith} and Rustlantis~\citep{Rustlantis-CHECKED} generate programs that find bugs in C compilers and in rustc, 
and \citet{CompilerTestingSurvey} survey compiler testing techniques. 
\citet{GCCandLLVM} study the bugs of GCC and LLVM, 
\citet{rustc-compiler-bugs-CHECKED} those of rustc, 
and \citet{CompilerImpactCommits} how developers work around compiler bugs. 
Closest to our work, 
\citet{Marcozzi-CHECKED} measured how fuzzer-found Clang and LLVM bugs affect Debian packages and found the impact rare. 
Our \rqtwo asks the same question of zkVM guests, where one miscompilation yields a valid proof of a wrong result, 
and our \rqthree asks what an attacker can do when the code is not fixed in advance. 
\citet{zkvm-optimisations} study how compiler optimisations change the performance of zkVM guests. 
We study how compiler bugs change their correctness.

\point{Security of ZKPs}
Prior work has studied vulnerabilities across the SNARK stack~\citep{sok-snarks-CHECKED}, 
surveyed zkVM designs~\citep{sok-understanding-zkvm-CHECKED}, 
and discussed what can go wrong in a zkVM~\citep{zksecurity-zkvm}. 
Most tools target hand-written circuits~\cite{zkp-security-state-CHECKED}. 
They find under-constrained circuits by static analysis and SMT solving~\citep{picus,zkap,agnostic-CHECKED}, 
certify circuits with refinement types~\citep{coda}, 
and fuzz circuits~\citep{fuzzing-zkvm,zkfuzz}, 
and \citet{zkp-security-state-CHECKED} evaluate these tools and survey their use. 
Closer to us, MTZK~\citep{mtzk} and Circuzz~\citep{circuzz} test the compilers that turn circuit languages into constraints, 
and Arguzz~\citep{arguzz} and \citet{polygon-zkevm-vetting} test zkVMs and zkEVMs for soundness and completeness bugs. 
All of these look for bugs in the circuit or in the zkVM. 
We assume both are correct and study the compiler that builds the guest.

\point{Malicious contributions}
The xz backdoor~\citep{xz-backdoor} showed that an attacker can spend years building trust in an open-source project before inserting malicious code, 
and the hypocrite-commits study~\citep{hypocrite-commits} reported that patches with hidden vulnerabilities can pass maintainer review. 
In both cases the malicious behaviour is in the source. 
In ours the source is correct, and the behaviour appears only after compilation.

\section{Conclusion}
\label{sec:conclusion}
We presented the first empirical study of whether known miscompilations undermine the guarantees of zkVMs. We collected \nDataset{} rustc and LLVM miscompilations with Rust reproducers, using an LLM agent to reproduce \nLlvmRust{} of \nLlvmCandidates{} LLVM bugs from Rust. More than half of them carry over to zkVMs: 48 reproduce on SP1 and 46 on Jolt, and in each case the zkVM would prove a wrong output. Popular guests are rarely affected by accident. In 447 of the 552 pairs of a guest and a bug that built, the buggy and the fixed compiler produced byte-identical guests, and the other 105 pairs differ only in layout or agreed on every input we ran. However, an attacker can add such code on purpose. For about \$40 per project, an LLM agent wrote patches that look like ordinary optimisations and expose a known bug in 12 of 366 pairs, across 11 of \nRQThreeProjects{} SP1 projects. In 4 of them the guest commits a wrong output, among them an Ethereum client that rejects a valid block, and in the other 8 it fails on a valid input.

A zkVM proof covers the binary that the compiler produced, so a compiler bug is invisible to anyone who verifies the zkVM or reviews the guest's source. Projects should treat their toolchain as part of the trusted base and track the known bugs it contains. Maintainers should also review outside contributions with this in mind, since a change that preserves the behaviour of the source can still trigger a compiler bug.

\bibliographystyle{ACM-Reference-Format}
\bibliography{references}

\appendix
\section{Survey of Projects with zkVM Guest Programs}
\label{app:ecosystem}
This appendix describes how we built the list of projects that \Cref{sec:zkvm-selection} uses. 

\point{Search}
We searched GitHub for \code{Cargo.toml} files that depend on the guest or SDK crate of a zkVM, and collected the repositories that GitHub lists as dependents of these crates.
The two steps gave 2,588 repositories, or 2,440 without forks.
Two of the crate names also belong to unrelated projects, and after checking each such dependency we kept 2,416 repositories.
To find the guest programs, we searched each repository for the zkVM's entry-point macro, and relied on the layout of the repository when a guest has no such macro.

\point{Classification}
We classified only the repositories that have at least 5 stars, or belong to an organisation and received a push since January 2025.
The others, mostly small personal repositories, are unclassified.
We labelled each of the selected repositories from its description and the matched files, and read the repository itself when these were not enough.
We identified and removed 34 repositories that were unrelated.
We used four labels, as follows.
\emph{Production} projects are deployed on a blockchain or usable without modifications.
\emph{Work-in-progress} projects are incomplete and couldn't be easily used.
\emph{Benchmarks} contain benchmark payloads and comparisons between zkVMs.
\emph{Tutorials} show how to use a zkVM, and include examples and templates.
The final list has \nProjects{} projects, of which 266 are classified as production and 163 target more than one zkVM.

\begin{figure}[t]
\centering
\begin{tikzpicture}
\begin{axis}[
  xbar, bar width=4pt, width=0.7\linewidth, height=5.6cm,
  xmin=0, xmax=1500, enlarge y limits=0.1,
  symbolic y coords={Pico,Nexus,ZisK,OpenVM,Jolt,SP1,RISC Zero},
  ytick=data, tick label style={font=\scriptsize}, label style={font=\scriptsize},
  xlabel={Open-source projects with guest programs},
  nodes near coords, nodes near coords style={font=\tiny},
  legend style={font=\scriptsize, at={(0.97,0.03)}, anchor=south east, draw=none},
  reverse legend, area legend]
\addplot[fill=compilerblue!35, draw=compilerblue] coordinates {(1314,RISC Zero) (1003,SP1) (90,Jolt) (66,OpenVM) (53,ZisK) (40,Nexus) (39,Pico)};
\addplot[fill=compilerblue, draw=compilerblue] coordinates {(108,RISC Zero) (149,SP1) (4,Jolt) (10,OpenVM) (12,ZisK) (1,Nexus) (13,Pico)};
\legend{All projects, Production}
\end{axis}
\end{tikzpicture}
\caption{Open-source projects with zkVM guest programs, for the seven zkVMs with the most projects (September 2026; \nProjects{} projects in total). A project that targets several zkVMs counts once for each.}
\Description{Horizontal bar chart. RISC Zero 1314 projects of which 108 production; SP1 1003 of which 149; Jolt 90 of which 4; OpenVM 66 of which 10; ZisK 53 of which 12; Nexus 40 of which 1; Pico 39 of which 13.}
\label{fig:ecosystem}
\end{figure}
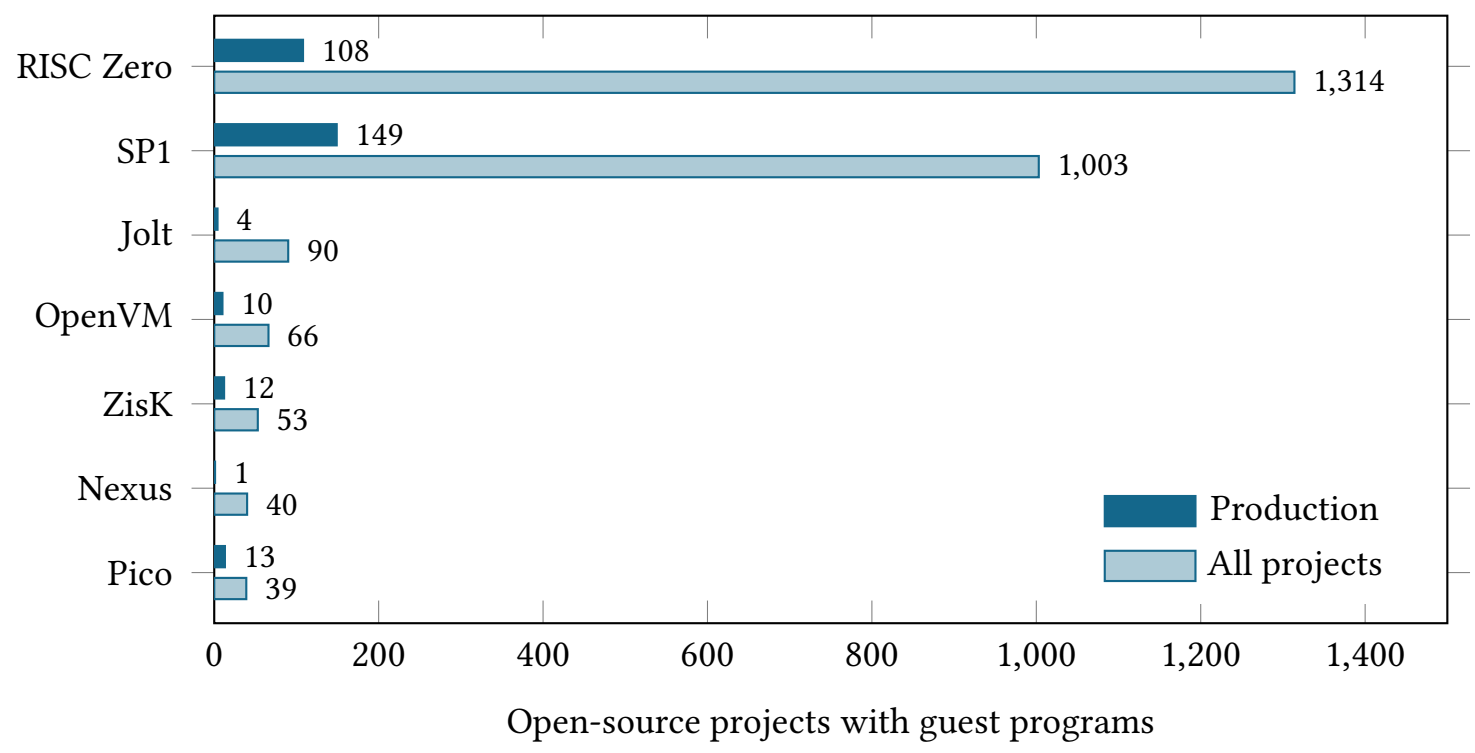

\section{Instructions Given to the LLVM-to-Rust Agent}
\label{app:llvm-to-rust-prompt}
We ran the LLVM-to-Rust agent in Codex with \code{gpt-5.6-sol} at high reasoning effort, once per LLVM bug and for at most four hours (\Cref{sec:bug-collection}).
Each run received a short launch prompt, which names the LLVM issue and points the agent to a protocol file in its workspace.
\Cref{lst:agent-prompt,lst:agent-protocol} give both in condensed form.
We keep the original wording, and {\color{gray}[\ldots]} marks text that we cut.
The cut text covers build parallelism, the timing of commands, the recording of notes, the layout of the saved files and the handling of retries.
We print the two tables of the protocol as lists.
The protocol defines eight outcomes, and every run ended in one of the first four, which are the outcomes of \Cref{tab:llvm-to-rust-outcomes}.
We keep only the bugs whose outcome is \code{rust-runtime-confirmed}.
The complete files are part of our artifact.

\begin{lstlisting}[style=prompt,caption={Launch prompt of an LLVM-to-Rust run (condensed).},label={lst:agent-prompt}]
Investigate LLVM issue <issue> and attempt a valid Rust reproducer of that known miscompilation.

Read protocol/experiment.md, protocol/result.schema.json, cases/<issue>/manifest.json,
cases/<issue>/progress.json, and cases/<issue>/history.jsonl. The protocol contains the complete
operating instructions, evidence requirements, outcome definitions, and output layout. [...]
Attempt the complete protocol in order, including compiler provisioning when required, until the
evidence gate or budget prevents progress.
[...]
Keep the Rust source revision constant when varying the known LLVM fix. [...] Compiler builds must
use the selected LLVM source rather than downloaded LLVM or rustc. [...] Confirm
semantic/behavioral LLVM failure before Rust searches. [...]

Use ordinary compilation for final acceptance. Prefer safe Rust and explicit arithmetic semantics.
Check expected outputs independently, outside compiled candidate assertions. A matched instruction
or changed assembly is insufficient. Screen generated candidates for the required IR/assembly
properties, then carry promising candidates through source-validity checks, an independent oracle,
buggy/fixed execution with unoptimised controls, and fresh replay in this same attempt. [...] do
not report runtime confirmation without completed end-to-end evidence.
[...]
\end{lstlisting}

\begin{lstlisting}[style=prompt,caption={Protocol followed by the LLVM-to-Rust agent (condensed).},label={lst:agent-protocol}]
# LLVM-to-Rust investigation protocol
[...]
## Workflow and evidence gates
(*\pkey{normalize:}*)        Join the candidate CSV, index, archived reports, patches, and all reproducers
(*\pkey{source-selection:}*) Find the introduction/fix history and compatible Rust/LLVM revisions
(*\pkey{llvm-baseline:}*)    Run the original LLVM case with and without its fix
(*\pkey{toolchains:}*)       Build the controlled rustc pair and compatible libraries
(*\pkey{rust-search:}*)      Search for valid Rust that naturally produces the necessary structure
(*\pkey{oracle:}*)           Establish expected behavior independently
(*\pkey{comparison:}*)       Compare reference and buggy/fixed compilations at the same settings
(*\pkey{replay:}*)           Minimise and repeat a successful experiment independently

[...] Prefer a naturally affected LLVM baseline plus the fix. If compatibility requires removing a
fix from a newer revision, label the pair as reconstructed; this does not establish that a
released Rust compiler was affected. Hold bootstrap, dependencies, target, configuration, and Rust
source constant across the pair. [...]

[...] Do not manually add LLVM IR or metadata to the final Rust program. Disabling passes or using
debug counters is diagnostic evidence and must be classified separately. A changed IR or assembly
file alone does not demonstrate incorrect behavior.
[...]
For every candidate, define the input domain and a reference with explicit bit widths. Check
overflow, shifts, initialization, pointer validity, aliasing, and intrinsic preconditions; justify
any unsafe Rust. Miri supplies evidence for tested executions, not universal validity. [...]

The comparison must show the independent expected result, buggy and fixed outputs at the same
relevant settings, and unoptimized controls for both builds. [...] Attribute the failure to the
known LLVM fix. Record any different defect separately. Minimize a successful Rust program and
input, then replay from saved artifacts in a fresh process.
[...]
## Checkpoints and classifications
[...] At the end of every attempt, report the strongest justified outcome:
(*\pkey{rust-runtime-confirmed:}*) Valid Rust returns an incorrect result under ordinary compilation, the
                        known LLVM fix restores the independent expected result, and a fresh
                        replay confirms it
(*\pkey{rust-ir-confirmed:}*)      Valid Rust reaches the relevant defective IR transformation, but runtime
                        evidence is incomplete
(*\pkey{diagnostic-only:}*)        Evidence depends on a modified pipeline, debug counter, or other
                        diagnostic option
(*\pkey{llvm-only-confirmed:}*)    The original LLVM defect is confirmed; a Rust reproduction remains
                        unresolved
(*\pkey{toolchain-blocked:}*)      The required compatible compiler combination could not be constructed
(*\pkey{invalid-candidate:}*)      The candidate violates source-language requirements
(*\pkey{budget-exhausted:}*)       The search budget ended without sufficient evidence for a stronger outcome
(*\pkey{unresolved:}*)             Available evidence supports none of the above outcomes

An unsuccessful search does not prove that Rust cannot generate the structure.
[...]
\end{lstlisting}

\section{Instructions Given to the Patch-Search Agent}
\label{app:rq3-prompt}
We ran the patch-search agent of \rqthree in Codex with \code{gpt-6-sol} at medium reasoning effort (\Cref{sec:rq3-method}).
Every run used the same task file.
The message that starts a run points the agent to this file and gives it one project, the zkVM, which is always SP1, and the \nSPRepro{} bugs that reproduce on SP1 in \rqone.
The agent chooses which of these bugs to try, and has four hours of wall-clock time for the project.
\Cref{lst:rq3-task} gives the task file in condensed form.
We keep the original wording, and {\color{gray}[\ldots]} marks text that we cut.
The cut text covers how the inputs are resolved, the records of commands, timings, tests and token usage that the agent must keep, the layout of the saved files and the final report.
We print the outcome table as a list, and we show the inputs with the values that our runs used.
The four outcomes and the patch rules are the ones of \Cref{sec:rq3-method}, where the \emph{normal} compiler arm is the fixed compiler.
The complete file is part of our artifact.

\begin{lstlisting}[style=prompt,caption={Task file of the patch-search agent (condensed).},label={lst:rq3-task}]
# Test compiler bugs against a zkVM project

You are investigating whether known Rust or LLVM compiler bugs can affect the guest programs of a
real zkVM project. Work autonomously, keep every experiment local, and preserve enough evidence for
another engineer to review or replay the result.

## Inputs
(*\pkey{BUGS\_PATH=}*)<the 33 LLVM bugs that reproduce on SP1>
(*\pkey{ZKVM=}*)sp1
(*\pkey{PROJECT=}*)<GitHub URL of one of the 20 selected projects>
(*\pkey{TIME\_BUDGET=}*)<4 hours of wall-clock time for the project>
[...]
## Experiment rules
[...] Do not commit to, push to, or open a pull request against the selected project.

Use ordinary Rust source and the project's normal zkVM build path. A candidate patch must be a
plausible refactoring or optimization of existing guest code. It must preserve the intended source
behavior. Examples include replacing an equivalent loop, removing a redundant copy, changing an
equivalent arithmetic form, specializing a helper, or using a more direct data representation.

Do not paste a standalone reproducer into an unrelated guest, add manually written LLVM IR, inject
compiler metadata, add undefined behavior, weaken an assertion, change a test's expected value, or
alter the input only to make the two compiler arms perform different computations. [...]

The patched guest source, dependencies, features, target, build profile, linker, environment, and
input must be identical in the normal and buggy compiler arms. Only the compiler construction
needed to expose the selected bug may differ. [...]

A bug is eligible for guest patching only after both compiler arms can compile a real guest through
the selected zkVM's ordinary guest build pipeline. Treat this as a hard preflight gate. [...]

Compiler instrumentation, pass disabling, debug counters, and custom diagnostic flags may screen
candidates. They cannot establish success. [...]

Execute guests only. Do not generate or verify cryptographic proofs. [...]

## Phase 1, select and build the project
[...] Identify the command that compiles each selected guest for the requested zkVM. [...]
## Phase 2, discover existing execution and testing paths
Find everything the project already uses to exercise the selected guests [...]
## Phase 3, establish a small execution baseline
Run a small subset of the paths found in Phase 2. Prefer the fastest deterministic command that
executes each guest with a fixed input. [...]
## Phase 4, investigate each selected bug
### Qualify the compiler pair
Provision the normal and buggy compilers before searching for project patch sites. Prefer two
builds from the same compiler source revision with only the known fix applied or reverted. [...]
### Search the guests and design candidate patches
Before proposing a patch, turn the corpus reproducer into a concrete trigger specification. [...]

Search guest source and project crates reachable from the guest for natural places where a
refactoring or optimization inspired by that reproducer can introduce the trigger. [...] Prioritize
code that affects observable guest output and can receive the reproducer's witness values through
valid project inputs.

Try at most 10 distinct candidate patches per bug. [...]
### Build and compare compiler arms
[...] For every candidate whose preliminary screen contains the required compiler input, build the
same patched guest in both arms. [...] Compare affected and fixed IR or assembly at the faulty
pass, then finalize the candidate's `pattern-hit`, `pattern-miss`, or `pattern-uncertain` status.
### Backpropagate the affected compiler state to guest input
Perform this step for every `pattern-hit`. [...] Start at the first instruction, flag, metadata
item, value, or control-flow edge that differs because of the bug. Do not start from an assumed
test input. Work backward [...] until each required condition is expressed in terms of the guest's
external input. [...] Then trace forward from the affected value to an observable result. [...]
### Use the existing project workload when directed testing does not diverge
[...] use the remaining budget on the most useful tests, executions, fuzz seeds, or small fuzzing
run found in Phase 2. Run identical commands and inputs in both compiler arms. [...]

## Outcome classification
Report exactly one outcome per bug:
(*\pkey{success:}*)              A plausible guest patch that preserves semantics reaches the selected bug
                      pattern, the normal and buggy compiler arms produce different observable
                      execution results for the same patched source and input, the normal compiler
                      preserves the unpatched result, and a clean replay repeats the differential.
(*\pkey{success-without-test:}*) A plausible guest patch preserves checked source behavior, compiles through
                      the ordinary zkVM pipeline in both arms, and has fix-specific
                      compiler-boundary evidence with status `pattern-hit`, but no observable
                      execution divergence was confirmed before the budget ended. [...]
(*\pkey{partial:}*)              The compiler-pair eligibility gate passed and at least one plausible guest
                      refactoring or optimization was compiled with both compiler arms and
                      preserved the checked baseline behavior under the normal compiler, but every
                      candidate was `pattern-miss` or `pattern-uncertain`.
(*\pkey{failure:}*)              The compiler-pair eligibility gate did not pass, no valid project patch was
                      produced, either compiler arm could not build a real zkVM guest, the project
                      could not be executed, the candidate changed intended semantics, or available
                      evidence supports none of the other outcomes.
[...]
\end{lstlisting}
\end{document}